\documentclass[usenatbib]{mn2e}
\usepackage{graphicx}
\usepackage{amssymb}

\begin{document}

\title[Spectral variations from an orbiting spot]
{Variation in the primary and reprocessed radiation from an orbiting
spot around a black hole}

\author[M.~Dov\v{c}iak, V.~Karas, G.~Matt and R.W.~Goosmann]
{M.~Dov\v{c}iak,$^{\!1}$~%\,\thanks{E-mail: dovciak@astro.cas.cz}
 V.~Karas,$^{1}$
 G.~Matt$^{2}$ and
 R.W.~Goosmann$^{1}$\\~\\
$^1$~Astronomical Institute, Academy of Sciences of the Czech Republic,
 Bo\v{c}n\'{\i}~II, CZ-140\,31~Prague, Czech Republic\\
$^2$~Dipartimento di Fisica, Universit\`a degli Studi ``Roma Tre'',
 Via della Vasca Navale 84, I-00146~Roma, Italy}

\date{Accepted ... 2007, Received ... 2007}
\pagerange{\pageref{firstpage}--\pageref{lastpage}}
\pubyear{2007}
\maketitle
\label{firstpage}

\begin{abstract}
We study light curves and spectra (equivalent widths of the iron line and
some other spectral characteristics) which arise by reflection on the surface
of an accretion disc, following its illumination by a primary off-axis source
--- an X-ray `flare', assumed to be a point-like source just above the
accretion disc resulting in a spot with radius $\Delta r/r \lesssim 1$.
We consider General Relativity effects (energy shifts,
light bending, time delays) near a rotating black hole, and we find them all
important, including the light bending and delay amplification due to the spot
motion. For some sets of parameters the observed reflected flux exceeds the
observed flux from the primary component.
We show that the orbit-induced variations of the equivalent width with
respect to its mean value can be as high as 30\% for an observer's inclination
of 30$^{\circ}$, and much more at higher inclinations. We calculate the
ratio of the reflected flux to the primary flux and the hardness ratio which
we find to vary significantly with the spot phase mainly for small orbital
radii. This offers the chance to estimate the lower limit of the black hole
spin if the flare arises close to the black hole.
\end{abstract}


\section{Introduction}
X-ray spectral measurements of the iron line and the underlying continuum
provide a powerful tool to study accretion discs in active galactic
nuclei (AGN) and Galactic black holes (for a review see Fabian et al.\
\citeyear{fab00}; Reynolds \& Nowak \citeyear{rey03}). If a line originates by
reflection of the primary continuum, then its observed characteristics
may reveal rapid orbital motion and light bending near the central black hole.
Spectral characteristics can be employed to constrain the black hole mass
and angular momentum. A particularly important role is played by the
equivalent width (EW), which reflects the intensity of the line versus the
continuum flux as well as the role of General Relativity effects in the
source. In order to reduce the ambiguity of results one needs to perform
spectral fitting with self-consistent models of both the line and
continuum.

Some AGN are known to exhibit EW greater than expected for a ``classical''
accretion disc. Enhanced values for the EW can be obtained by assuming
an anisotropical distribution of the primary X-rays
(Ghisellini et al.\ \citeyear{ghi91}),
significant ionization of the disc matter (Matt, Fabian \& Ross
\citeyear{mat93})
or iron overabundance (George \& Fabian \citeyear{geo91}).
Martocchia \& Matt (\citeyear{mar96}) and Martocchia et al.\/
(\citeyear{mar00}) found, using an axisymmetric lamp-post
scheme, an anticorrelation between the intensity of the reflection features and
the primary flux. When the primary source is at a low height on the disc axis,
the EW can be increased by up to an order of magnitude with respect
to calculations neglecting General Relativity effects.
When allowing the source to be located off the axis of rotation, an
even stronger enhancement is expected (Dabrowski \& Lasenby \citeyear{dab01}).
Miniutti et al.\/ (\citeyear{min03}) and Miniutti \& Fabian (\citeyear{min04})
have realised that this so-called light
bending model can naturally explain the puzzling behaviour of the iron line of
MCG--6-30-15, when the line saturates at a certain flux level and then
its EW starts decreasing as the continuum flux increases further.
Clear understanding of the interplay between the primary and the reprocessed
components is therefore highly desirable.

In our previous paper (Dov\v{c}iak et al.\ 2004a), we have proposed that
the orbiting spot model could explain the origin of transient narrow
lines, which have been reported in some AGN X-ray spectra (Turner et al.\ 2002;
Guainazzi 2003; Yaqoob et al.\ 2003) and widely discussed since then. The main
purpose of the current paper is to present accurate
computations of time-dependent EWs and other spectral characteristics within
the framework of the spot model, taking into account a consistent scheme for
the local spectrum reprocessing. The main difference from previous
papers is that the current one connects the primary source power-law
continuum with the reprocessed spectral features. Both components are
further modified by relativistic effects as the signal propagates
towards an observer.

In Section \ref{model} we describe the model and we summarize the equations
and the approximation used. In Section \ref{results} we present the results
of our calculations. The final conclusions are drawn in Section
\ref{conclusions}.

\section{Model set-up and equations}
\label{model}

\subsection{Model approximations and limitations}

\begin{figure}
\begin{center}
\includegraphics[width=7cm]{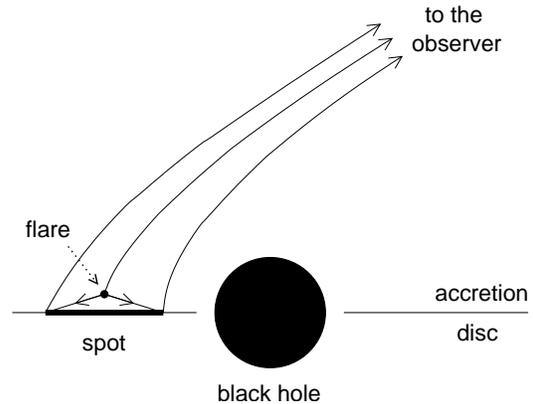}
\end{center}
\caption{A sketch of the model geometry (not to scale).
A localized flare occurs above the
disc, possibly due to magnetic reconnection, and creates a spot
by illuminating the disc surface. The resulting `hot spot' co-rotates with the
disc and contributes to the final observed signal by reprocessing
the primary X-rays.}
\label{fig-model}
\end{figure}

We examine a system composed by a black hole, an accretion disc and a
co-rotating flare with the spot underneath (Collin et al.\ \citeyear{col03};
see Fig.~\ref{fig-model}). The gravitational field is described in terms of
Kerr metric (Misner, Thorne \& Wheeler \citeyear{mis73}). Both static
Schwarzschild and rotating Kerr black holes are considered. The
co-rotating Keplerian accretion disc is geometrically thin and optically
thick, therefore we take into account only photons coming from the
equatorial plane directly to the observer. We further assume that the
matter in the accretion disc is cold and neutral.

A flare is supposed to arise in the disc corona due to a magnetic
reconnection event (e.g.\ Galeev, Rosner \& Vaiana \citeyear{gal79};
Poutanen \& Fabian \citeyear{pou99};
Merloni \& Fabian \citeyear{mer01}; Czerny et al.\ \citeyear{cze04}). Details
of the formation of the flare and its
structure are not the subject of the present paper instead we assume
that the flare is an isotropic stationary point source with a power-law
spectrum. It is located very near above the disc surface and
it co-rotates with
the accretion disc. We also assume that the single flare dominates the
intrinsic emission for a certain period of time.

The question of the flare height above the accretion disc is still
unresolved. Although there are some similarities between the
reconnection events that are responsible for solar flares and those in
accretion discs (Romanova et al. 1998; Czerny et al.\ \citeyear{cze04}),
it is not
yet clear to what height the magnetic loops can rise in the latter case.
For example, Dabrowski \& Lasenby (2001) {\em assumed} a large
height ($h(r)\simeq1$--$2R_{\rm{}g}$) of the flare above the inner disc region
and they performed the ray tracing from the flare towards the disc surface and
further to the observer. On
the other hand, here we assume that the flare height is less than the
gravitational radius, which seems to be substantiated by the condition of
equipartition between the magnetic and gas pressure, but the
confirmation will need coupled radiation-magnetohydrodynamic
computations, which have been only recently started (e.g. Blaes et al.
2006). The small height allows us to simplify the calculations of the disc
irradiation by neglecting the light-bending on primary rays, so that we
can study off-axis flares and manage the time-dependent evolution of the
observed flux, EW and other characteristics.

The spot represents the flare-illuminated part of the disc surface.
The flare can be viewed as a lamp. The reprocessed photons are re-emitted from
only those parts of the disc that are illuminated by the flare.
Therefore the spot does not share the differential rotation with the disc
material. It is considered to be a rigid two dimensional circular feature,
 with its
centre directly below the flare. However, the matter in the disc
lit by the flare is in Keplerian orbits at the corresponding radii, thus
it has different velocities at different parts of the spot. This is
important when calculating the transfer function for the observer in the
infinity. Because the flare is very close to the disc, the spot does not
extend far from below the flare. Only photons emitted into the lower
hemisphere reach the disc. We assume the opening angle of the
illuminating cone to be $89^\circ$ which means we lose less than 2\% of these
photons. We neglect the photons emitted into the upper hemisphere directed
towards the black hole that could possibly reach the disc behind
it. We approximate the photon trajectories between the flare and
the spot by straight lines. We do not consider the energy shift and
abberation due to the different motion of the flare and the illuminated
disc matter. As far as the illumination is concerned we consider the
flare and the spot to be in the same co-moving reference frame.
Thus the illumination of the spot is approximated by a simple cosine law in
the local Keplerian frame co-moving with the matter in the spot and decreasing
with the distance from the flare (i.e.\ illuminating flux
$\sim \mu_{\rm i}^3/h^2$, where $\mu_{\rm i}$ is the cosine of the incident
angle in the local Keplerian frame and $h$ is the height of the flare above the
disc).
Furthermore we neglect the time delay between the photon's emission from the
flare and its later re-emission from the spot.

The intrinsic (local) spectra from the spot were computed by Monte Carlo
simulations considering multiple Compton scattering and iron line fluorescence
in a cold, neutral, constant density slab with solar iron abundance. We used
the NOAR code for these computations (see Section 5 of Dumont et al.\/
\citeyear{dum00} and Chapter 5 of Goosmann \citeyear{goo06}).
The local flux depends on the local incident and local emission angles, hence
the flux changes across the spot. Here and elsewhere in the text we
refer to the quantities measured in the local frame co-moving with the matter
in the disc as ``local''.

The local flux consists of only two components --- the flux from the
primary source (the flare) and the reflected flux from the spot. The latter one
consists of the reflection continuum (with the Compton hump and the iron edge
as the main features) and the neutral K$\alpha$ and K$\beta$ iron lines. No
other emission is taken into account.

As far as the photon trajectories from the spot to the observer are concerned,
all general relativistic effects --- energy shift, aberration, light bending,
lensing and relative time delays --- are taken into account. We assume that
only the gravity of the central black hole influences photons on their travel
from the disc to the observer. This allows us to define a relatively simple
scheme in which different intervening effects remain under full control and can
be well identified.

\subsection{Model predictions for the observed flux}
\label{math}

The observed energy flux from the finite size spot on the accretion disc, $F$,
can be computed from the
local energy flux, $f$ (Dov\v{c}iak \citeyear{dov04a};
Dov\v{c}iak et.~al.\/ \citeyear{dov04c}),
\begin{equation}
\label{flux1}
F(E,t)=\int_{\Sigma}{\rm d}S\,f(E/g;r,\varphi;t_{\rm e})\,G(r,\varphi),
\end{equation}
where ${\rm d}S= r\,{\rm d}r\,{\rm d}\varphi$ is an area element on the disc,
\begin{equation}
\label{time}
t_{\rm e}(t,r,\varphi)=t-\delta t(r,\varphi)
\end{equation}
is the emission time, $\delta t$ is a relative time delay with
which photons emitted at different places on the disc reach the observer
(we use Boyer-Lindquist spheroidal coordinates;
Misner et al.\ \citeyear{mis73}).
Note that we use the specific energy flux
(not the photon flux) per unit solid angle (not per unit area),
i.e.\/ $F\equiv E\,{\rm d}N(E)/({\rm d}t\,{\rm d}E\,{\rm d}\Omega)$.
Hence the units are keV/s/keV for spectra (keV/s for integrated
energy flux in light curves). One can get the
flux per unit area of the detector by dividing our results by the distance of
the source squared.

In a classical (non-relativistic) case, when there is no light bending
and aberration, the relative time delay would be
$\delta t(r,\varphi)=r\,\sin{\theta}\sin{\varphi}$ (the observer is located in
the direction $\varphi=-90^\circ$).
The transfer function (see Cunningham \citeyear{cun75}) for an extended source
is $G = g^2\,l\,\mu_{\rm e}$, where $g$
denotes the combined gravitational and Doppler shift, $l$ is lensing and
$\mu_{\rm e}$ is the cosine of the local emission angle at the disc (measured
with respect to
the frame co-moving with the disc).
All these functions depend on the place of the emission
($r$ and $\varphi$ on the disc). Because the local flux is coming from an
orbiting spot it is useful to use coordinates co-moving with the spot's
centre. The observed flux is then
\begin{equation}
\label{flux2}
F(E,t)=\int_{\Sigma}{\rm d}S_{\rm e}\,f(E/g;r,\varphi;t_{\rm e})\,
G(r,\varphi)\,k_{\rm t}(r,\varphi),
\end{equation}
where ${\rm d}S_{\rm e}=r\,{\rm d}r\,{\rm d}\varphi_{\rm e}$
with
\begin{equation}
\label{phi}
 \varphi_{\rm e}=\varphi-\omega_{\rm e}\,[t-\delta t(r,\varphi)].
 \end{equation}
This equation defines $\varphi$ in eq.~(\ref{flux2}) as a function of
$\varphi_{\rm e}$ and $t$. The angular velocity $\omega_{\rm e}$ is the
angular velocity of the spot. The factor $k_{\rm t}$ arises from the coordinate
transformation $\varphi \rightarrow \varphi_{\rm e}$
(it is connected with the time delays). This is due to the spot moving in its
orbit towards or away from the observer and due to the fact that the light rays
emitted from different places on the disc are differently bent, thus acquiring
different travel times. Therefore we will refer to this factor as the delay
amplification,
\begin{equation}
\label{kt}
k_{\rm t}(r,\varphi) = \left[1+\omega_{\rm e}
\frac{\partial(\delta t)}{\partial \varphi}(r,\varphi)\right]^{-1}\ .
\end{equation}

Similarly as for the spot emission one gets an
expression for the observed energy flux from the primary source, the flare,
which we suppose is a point-like source
\begin{eqnarray}
\label{flux3}
F_{\rm p}(E,t)=f(E/g;t_{\rm e})\,G_{\rm p}(r,\varphi)\,k_{\rm t}(r,\varphi).
\end{eqnarray}
Here, $G_{\rm p} = g^3\,l$ is the transfer function for a point-like source and
the emission time $t_{\rm e}$ together with the azimuthal coordinate $\varphi$
are both functions of the observer's time $t$ as defined in
eqs.~(\ref{time}) and (\ref{phi}) with $\varphi_{\rm e}=0$.

\subsection{The intrinsic emission of the flare and the spot}

We define the primary emission to be
\begin{equation}
\label{primary_flux}
f_{\rm p}(E)\equiv E\,\frac{{\rm d}N_{\rm p}(E)}{{\rm d}t\,{\rm d}E\,
  {\rm d}\Omega_{\rm p}}= E^{1-\Gamma},
\end{equation}
where $\Gamma$ is a photon number density power-law index.

The energy flux of reflected photons is
\begin{equation}
\label{reflected_flux}
f_{\rm r}(E,\mu_{\rm i},\mu_{\rm e},\Phi_{\rm e}-\Phi_{\rm i})
  \equiv \frac{E\;{\rm d}N_{\rm r}(E)}{{\rm d}t\,{\rm d}E\,{\rm d}S_{\perp}
  {\rm d}\Omega_{\rm e}}= n_{\rm r}\,E\,\frac{\mu_{\rm i}^3}{h^2}
  \frac{1}{\mu_{\rm e}}\, ,
\end{equation}
where
\begin{equation}
\label{reflected_flux2}
  n_{\rm r}(E,\mu_{\rm i},\mu_{\rm e},\Phi_{\rm e}-\Phi_{\rm i}) =
  \frac{{\rm d}N_{\rm r}(E)}{{\rm d}t\,{\rm d}E\,
  {\rm d}\Omega_{\rm p}{\rm d}\Omega_{\rm e}}\ .
\end{equation}
is the photon number density flux of the reflected radiation emitted
into the solid angle ${\rm d}\Omega_{\rm e}$ if the incident light rays come
from the solid angle ${\rm d}\Omega_{\rm p}$. In eq.~(\ref{reflected_flux})
we used the fact that the solid angle ${\rm d}\Omega_{\rm p}$ corresponds to
the area ${\rm d}S_{\perp}$ perpendicular to
the light ray emitted from the spot in the following way:
\begin{equation}
\label{dS}
{\rm d}S_{\perp}=\frac{{\rm d}S_{\perp}}{{\rm d}S}
        \frac{{\rm d}S}{{\rm d}S_{{\rm i}\perp}}
        \frac{{\rm d}S_{{\rm i}\perp}}{{\rm d}\Omega_{\rm p}}\,
        {\rm d}\Omega_{\rm p}
       =\mu_{\rm e}\,\frac{h^2}{\mu_{\rm i}^3}\,{\rm d}\Omega_{\rm p}.
\end{equation}
Here, $h$ is the height of the primary source above the disc, ${\rm d}S$ is
the area of the spot lit by the emission coming from the solid angle
${\rm d}\Omega_{\rm p}$, ${\rm d}S_{{\rm i}\perp}$ is the corresponding area
perpendicular to the incident light ray and $\mu_{\rm i}$ is the cosine of the
incident angle. All these quantities are evaluated in the local reference
frame co-moving with the disc.

As mentioned earlier, the emitted photon flux $n_{\rm r}$ was calculated
by Monte Carlo simulations. In our computations we used pre-calculated
tables of $n_{\rm r}(E,\mu_{\rm i},\mu_{\rm e})$ which were averaged over the
difference between incident and emitted azimuthal angles
$\Phi=\Phi_{\rm e}-\Phi_{\rm i}$.

\subsection{The equivalent width, ratio of reflected and primary fluxes,
            hardness ratio}

\begin{figure}
\begin{center}
  \includegraphics[width=4.cm]{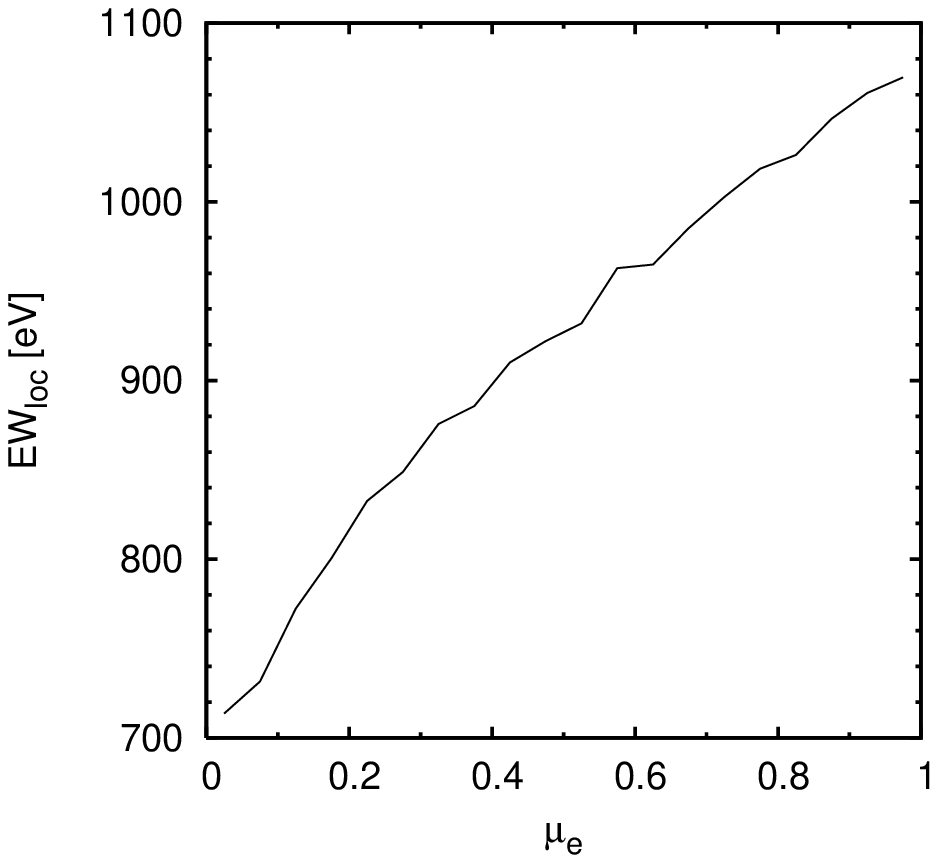}
  \includegraphics[width=4.cm]{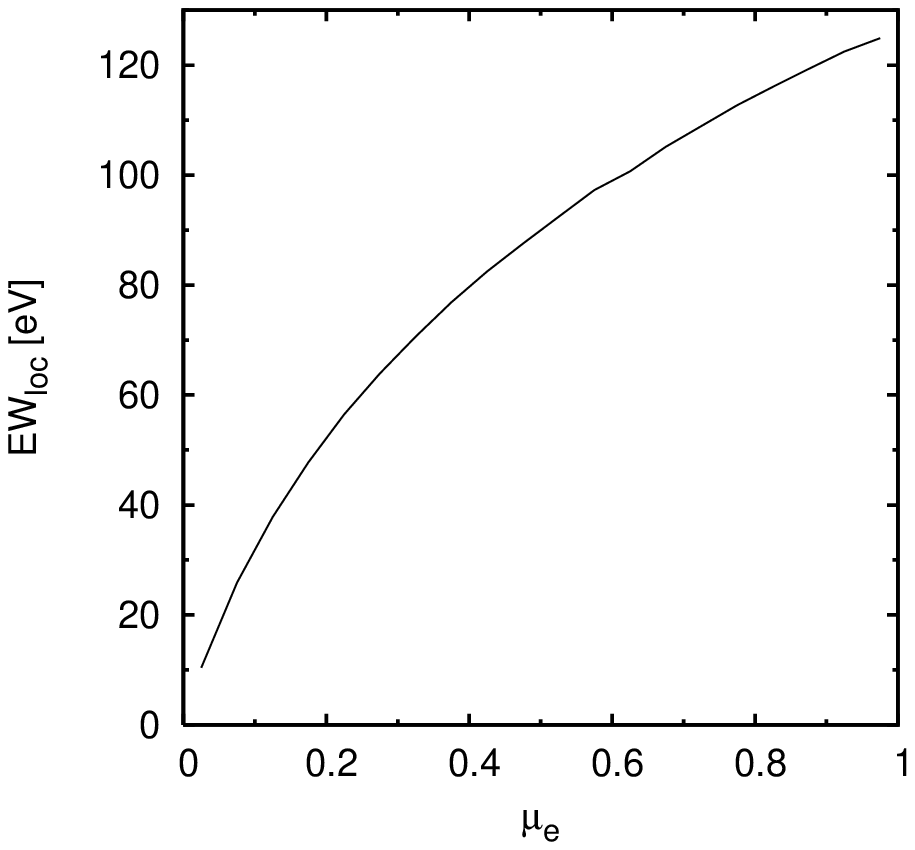}
  \caption{{\bf Left:} The local equivalent width without taking the primary
  flux   into account as a function of the direction of emission.
  {\bf Right:} The same as in the left panel but with the flux from the primary
  source included.}
  \label{fig-ew_loc}
\end{center}
\end{figure}

Because the flare is very near above the disc the spot receives the light
emitted downward to almost the whole half-space. The local equivalent width
of the spectral line is
\begin{eqnarray}
\nonumber
{\rm EW}_{\rm loc}(\mu_{\rm e}) & = &
 \frac{\int{\rm d}S \mu_{\rm e}\int{\rm d}E\,f_{\rm r}^{\rm L}(E)}
 {\int{\rm d}S \mu_{\rm e}f_{\rm r}^{\rm C}(E_{\rm L})+f_{\rm p}(E_{\rm L})}\\
\label{ew1}
& = & \frac{\int{\rm d}E\,\bar{n}_{\rm r}^{\rm L}(E,\mu_{\rm e})\,E}
{\bar{n}_{\rm r}^{\rm C}(E_{\rm L},\mu_{\rm e})E_{\rm L}+
E_{\rm L}^{1-\Gamma}/2\pi}\, ,
\end{eqnarray}
where we defined an average of both the line ($n_{\rm r}^{\rm L}$) and
continuum
($n_{\rm r}^{\rm C}$) part of the local photon flux emitted by a point source
into the half-space as
\begin{eqnarray}
\nonumber
\bar{n}_{\rm r}(E,\mu_{\rm e}) & \equiv & \frac{1}{2\pi}
\int_0^{2\pi}{\rm d}\Omega_{\rm p}\,n_{\rm r}(E,\mu_{\rm i},\mu_{\rm e},\Phi)\\
\label{average_flux}
& = & \frac{1}{2\pi}\int_0^{1}{\rm d}\mu_{\rm i}\int_0^{2\pi}{\rm d}\Phi\,
n_{\rm r}(E,\mu_{\rm i},\mu_{\rm e},\Phi)\, .
\end{eqnarray}
The dependence of the local equivalent width on the emission angle is shown in
Fig.~\ref{fig-ew_loc}.

\begin{figure}
\begin{center}
  \includegraphics[width=4.cm]{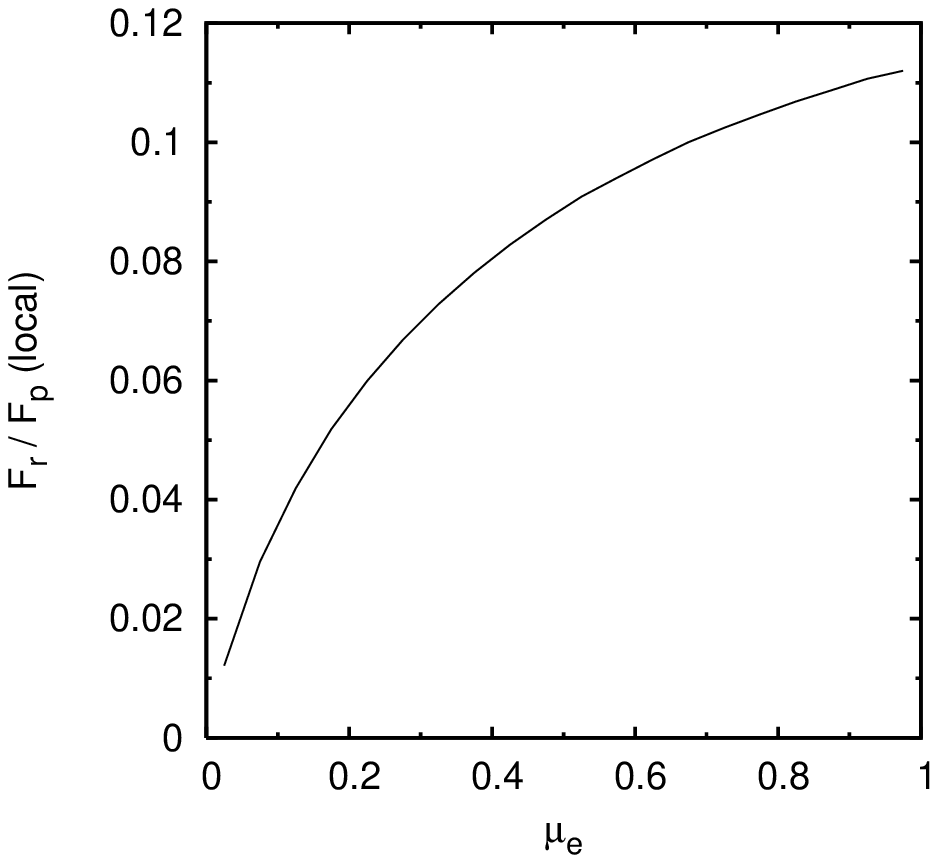}
  \includegraphics[width=4.1cm]{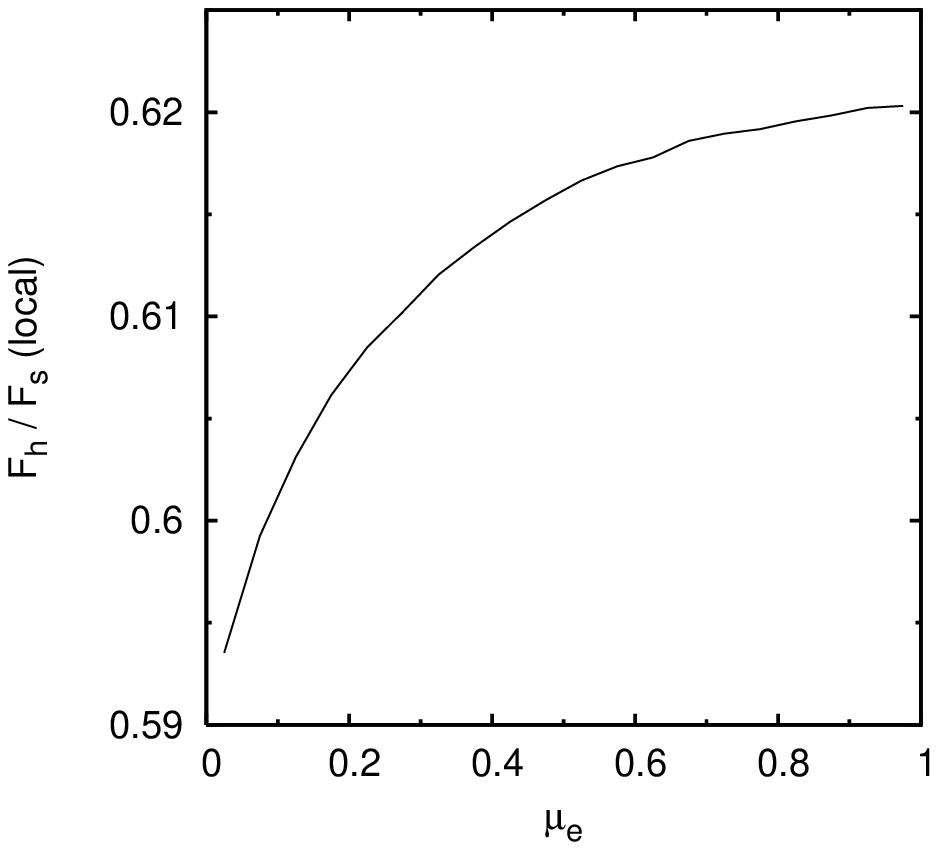}
  \caption{{\bf Left:} The ratio of the locally emitted energy flux in the
  direction   $\mu_{\rm e}$ to the primary flux. The fluxes are integrated in
  the energy   range 3--10~keV.
  {\bf Right:} The local hardness ratio of the fluxes in the ranges
  6.5--10~keV ($F_{\rm h}$) and 3--6.5~keV ($F_{\rm s}$).}
  \label{fig-ratio_loc}
\end{center}
\end{figure}

In the relativistic case the equivalent width is
\begin{eqnarray}
\nonumber
{\rm EW}(t)\hspace*{-1mm} & = & \hspace*{-1mm}
\frac{\int{\rm d}S_{\rm e}\, Gk_{\rm t}g
\int{\rm d}E\,f_{\rm r}^{\rm L}(E)}
{\int{\rm d}S_{\rm e}\, Gk_{\rm t}\,f_{\rm r}^{\rm C}(E_{\rm c}/g)+
G_{\rm p}^{\rm s} k_{\rm t}^{\rm s}\,f_{\rm p}(E_{\rm c}/g_{\rm s})}\\
\nonumber
& \approx & \hspace*{-1mm}\frac{g\int{\rm d}E\,\bar{n}_{\rm r}^{\rm L}
(E,\mu_{\rm e})\,E}{\bar{n}_{\rm r}^{\rm C}(E_{\rm L},\mu_{\rm e})E_{\rm L}+
E_{\rm L}^{1-\Gamma}/2\pi}\\[1mm]
\label{ew2}
& = & \hspace*{-1mm}
g(t)\; {\rm EW}_{\rm loc}(\mu_{\rm e}(t))\, .
\end{eqnarray}
A factor $g$ in the numerator accounts for integration of the line flux
over local energy. Note that the amplification function for the primary flux is
taken at the spot centre (denoted by index `s'), whereas this
function for the line and reflected continuum emission is changing throughout
the spot. In the transformation from coordinate area ${\rm d}S_{\rm e}$ to the
local solid angle ${\rm d}\Omega_{\rm p}$ we used the fact that the local area
will differ from the coordinate one by the $g$-factor.
The centroid energy in eq.~(\ref{ew2}) is defined as
\begin{equation}
\label{centroidE}
E_{\rm c}(t)=\frac{\int{\rm d}E\, F_{\rm r}^{\rm L}(E) E}{\int{\rm d}E\,
F_{\rm r}^{\rm L}(E)}
\approx g(t) E_{\rm L}\,
\end{equation}
with $F_{\rm r}^{\rm L}(E)$ being the observed energy flux
in the line.
The approximations in eqs.~(\ref{ew2}) and (\ref{centroidE}) hold only
for a very small ``effective'' spot (by which we mean
a patch of the disc from which photons arrive simultaneously; the observed spot
shape is deformed by the time delays).
In that case the energy shift $g$ and the overall amplification $Gk_{\rm t}$
change only
slightly throughout the spot and can be represented by their values at the
centre of the spot, $g_{\rm s}$ and $G^{\rm s}k_{\rm t}^{\rm s}$, respectively.

The local ratio of the reflected emission to the primary
radiation within the energy range $\langle E_1, E_2\rangle$ is
\begin{eqnarray}
\nonumber
\left  .\frac{F_{\rm r}}{F_{\rm p}}\right |_{\rm loc}
(E_1,E_2,\mu_{\rm e})\hspace*{-1mm} & = \hspace*{-1mm} &
\frac{\int{\rm d}S\mu_{\rm e}\int_{E_1}^{E_2}{\rm d}E\, f_{\rm r}(E)}
{\int_{E_1}^{E_2}{\rm d}E\, f_{\rm p}(E)}\\
\label{local_flux_ratio}
\hspace*{-1mm} & = \hspace*{-1mm} &
\frac{2\pi(2-\Gamma)\int_{E_1}^{E_2}{\rm d}E\,
\bar{n}_{\rm r}(E,\mu_{\rm e})E}{E_2^{2-\Gamma}-E_1^{2-\Gamma}}\, .
\end{eqnarray}
The dependence of the ratio of the reflected and primary radiation
on the emission angle is shown in the Fig.~\ref{fig-ratio_loc}.
In the relativistic case we find
\begin{eqnarray}
\nonumber
\frac{F_{\rm r}}{F_{\rm p}}(E_1,E_2,t)
\hspace*{-1mm} & =  & \hspace*{-1mm}
\frac{\int{\rm d}S_{\rm e}Gk_{\rm t}\int_{E_1}^{E_2}{\rm d}E\,
f_{\rm r}(E/g)}{G_{\rm p}^{\rm s}k_{\rm t}^{\rm s}
\int_{E_1}^{E_2}{\rm d}E\, f_{\rm p}(E/g_{\rm s})}\\
\nonumber
& \approx & \hspace*{-1mm}
\frac{2\pi(2-\Gamma)\int_{E_1}^{E_2}{\rm d}E\,
\bar{n}_{\rm r}(E/g,\mu_{\rm e})E}
{g^\Gamma(E_2^{2-\Gamma}-E_1^{2-\Gamma})}\\
\label{flux_ratio}
& = & \hspace*{-1mm}
\frac{k_{\rm r}(E_1,E_2,t)}{g(t)^\Gamma}\left .
\frac{F_{\rm r}}{F_{\rm p}}\right |_{\rm loc}(E_1,E_2,\mu_{\rm e}(t))
\end{eqnarray}
with the coefficient $k_{\rm r}(E_1,E_2,t)$ being
\begin{equation}
\label{coefficient}
k_{\rm r}(E_1,E_2,t) \equiv \frac{\int_{E_1}^{E_2}{\rm d}E\,
\bar{n}_{\rm r}(E/g(t),\mu_{\rm e}(t))E}
{\int_{E_1}^{E_2}{\rm d}E\,\bar{n}_{\rm r}(E,\mu_{\rm e}(t))E}\, .
\end{equation}
Again, the approximation in eq.~(\ref{flux_ratio}) holds true only when the
spot is small.

The last property we will discuss is the hardness ratio. The dependence
of the local hardness ratio, i.e.\/ the ratio of the hard flux $F_{\rm h}$
in the energy range $\langle E_2,E_3\rangle$ to the soft flux $F_{\rm s}$
in the energy range $\langle E_1,E_2 \rangle$,
\begin{equation}
\label{local_hardness_ratio}
\left .\frac{F_{\rm h}}{F_{\rm s}}\right |_{\rm loc}(\mu_{\rm e}) =
k\, \frac{\left .\frac{F_{\rm r}}{F_{\rm p}}\right |_{\rm loc}
(E_2,E_3,\mu_{\rm e})+1}
{\left .\frac{F_{\rm r}}{F_{\rm p}}\right |_{\rm loc}
(E_1,E_2,\mu_{\rm e})+1}\, ,
\end{equation}
on the emission angle is shown in Fig.~\ref{fig-ratio_loc}. Note that the
primary and both components of the reflected spectrum, line as well as
continuum, are taken into account. The $k$ factor is defined as the ratio of
the hard and soft primary fluxes,
\begin{equation}
\label{k-factor}
k=\frac{F_{\rm p}(E_2,E_3)}{F_{\rm p}(E_1,E_2)}.
\end{equation}
In the relativistic case we get
\begin{eqnarray}
\nonumber
\frac{F_{\rm h}}{F_{\rm s}}(t) & = &
k\, \frac{\frac{F_{\rm r}}{F_{\rm p}}(E_2,E_3,t)+1}
{\frac{F_{\rm r}}{F_{\rm p}}(E_1,E_2,t)+1}\\[1mm]
\label{hardness_ratio}
& \hspace*{-2cm} \approx & \hspace{-1.1cm}
k\, \frac{k_{\rm r}(E_2,E_3,t)\, g(t)^{-\Gamma}\left .
\frac{F_{\rm r}}{F_{\rm p}}\right |_{\rm loc}(E_2,E_3,\mu_{\rm e}(t))+1}
{k_{\rm r}(E_1,E_2,t)\, g(t)^{-\Gamma}\left .
\frac{F_{\rm r}}{F_{\rm p}}\right |_{\rm loc}(E_1,E_2,\mu_{\rm e}(t))+1}.
\end{eqnarray}

\section{Results}
\label{results}

\begin{figure*}
  \vspace*{-2mm}
  \includegraphics[height=4.7cm]{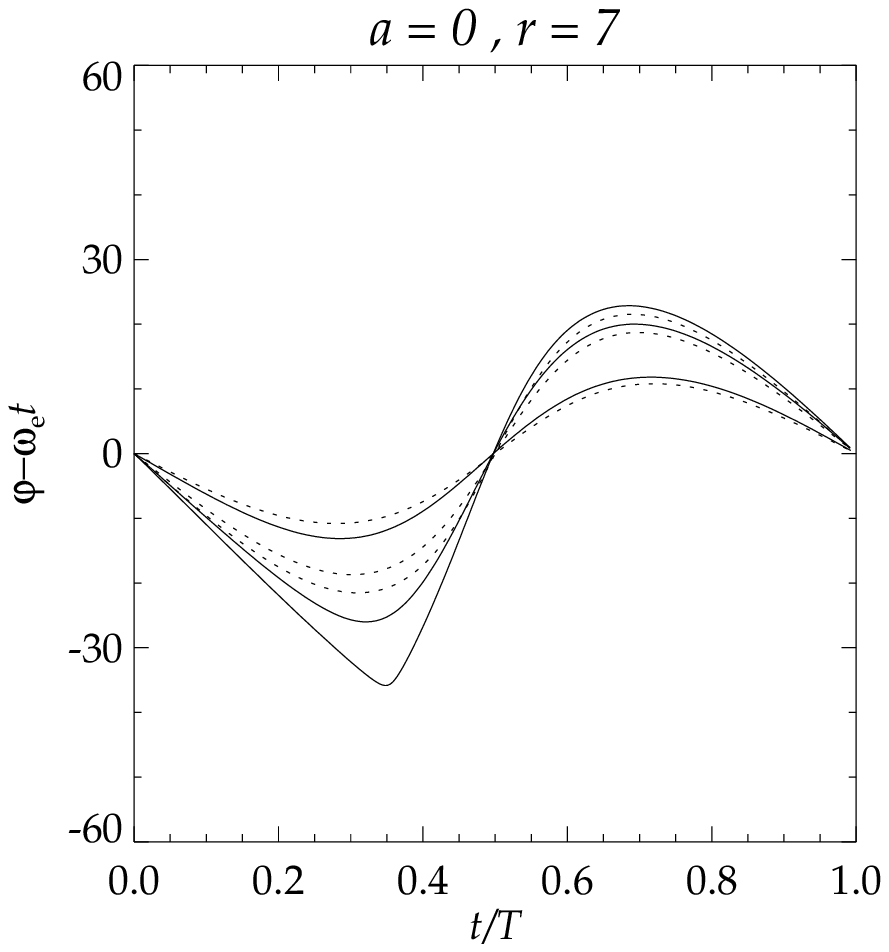}
  \hfill
  \includegraphics[height=4.7cm]{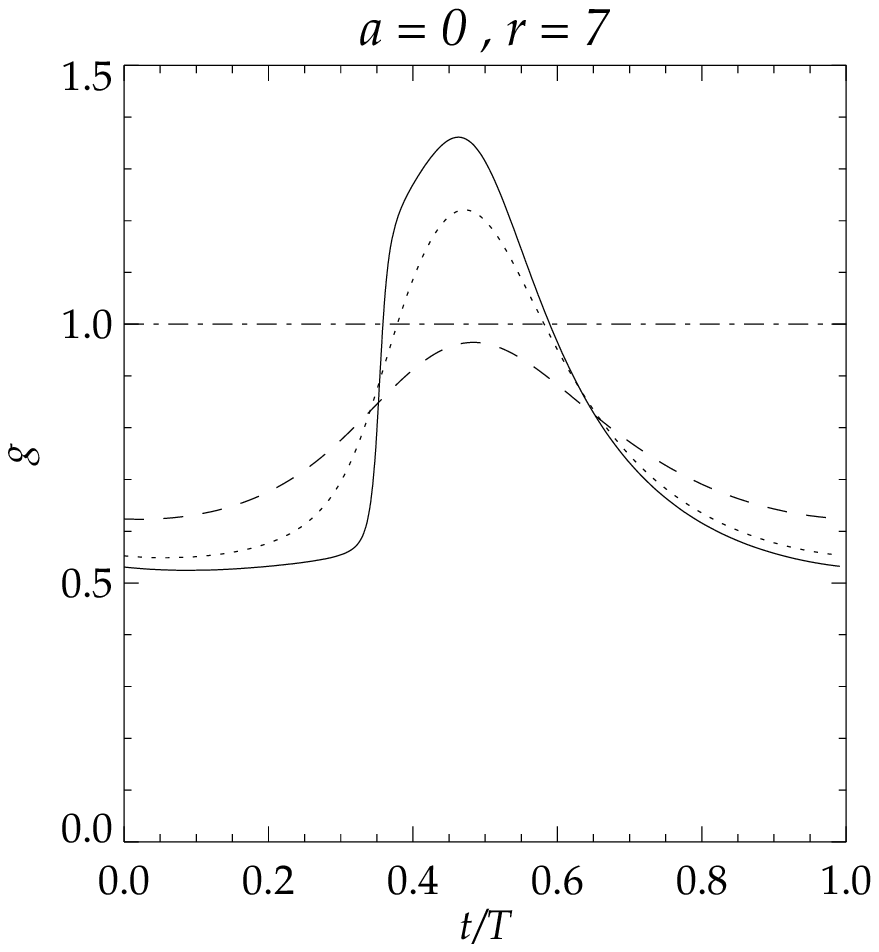}
  \hfill
  \includegraphics[height=4.7cm]{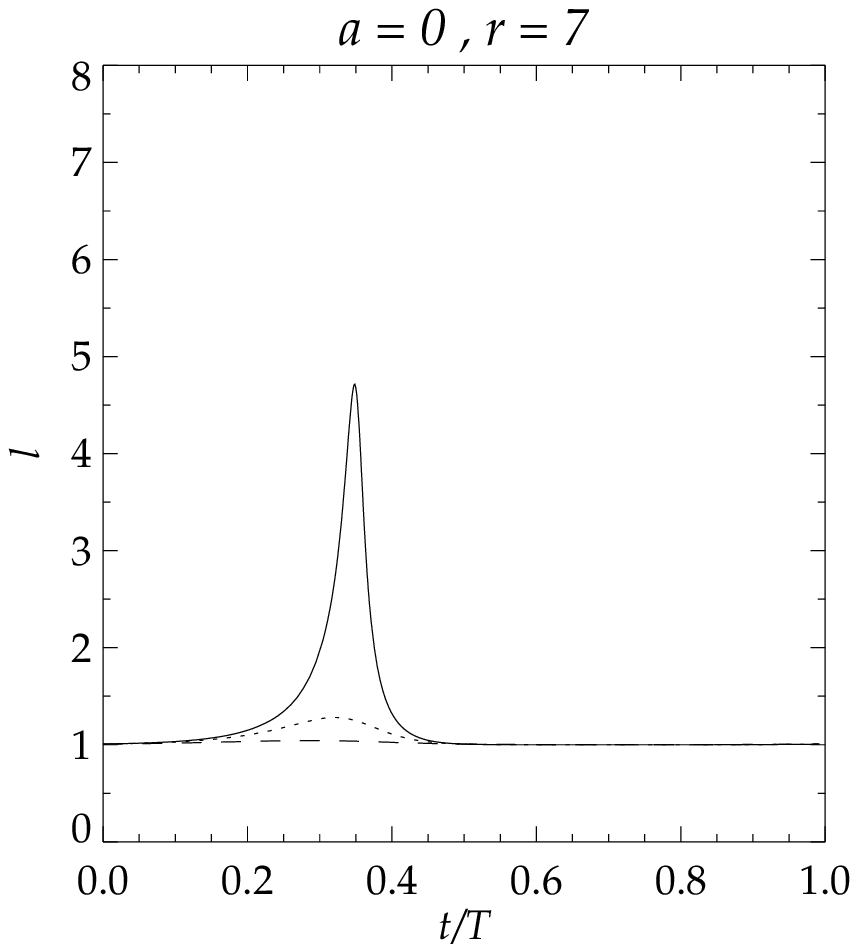}
  \hfill
  \includegraphics[height=4.7cm]{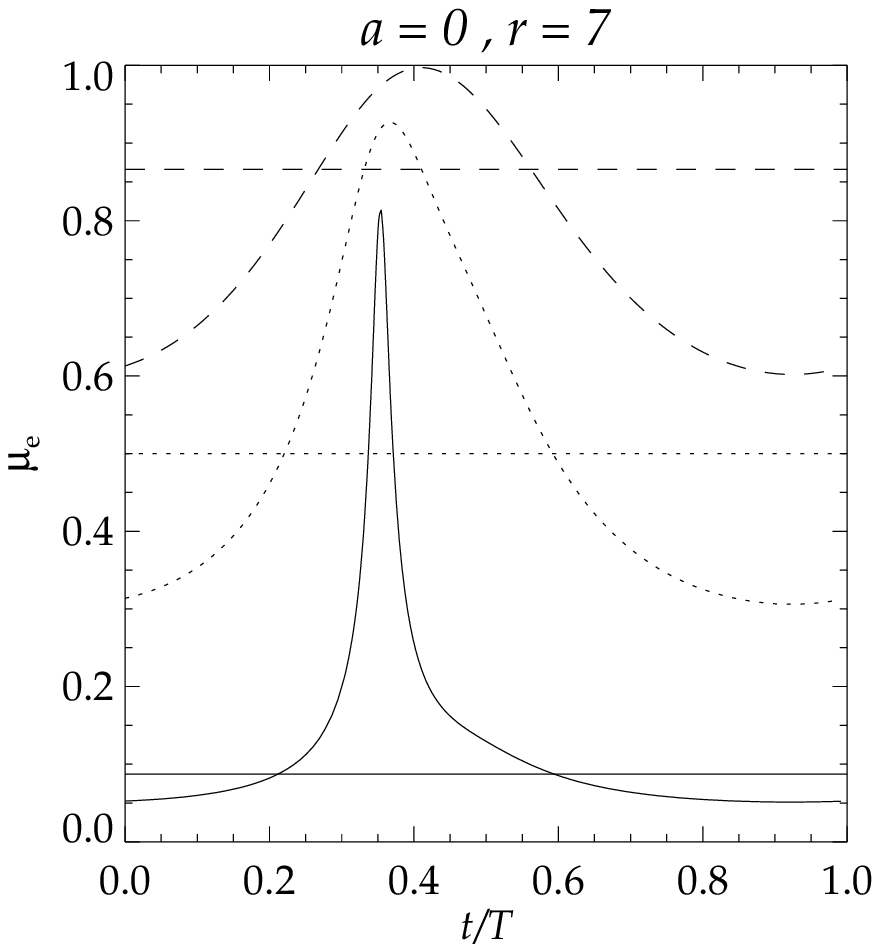}\\
  \includegraphics[height=4.7cm]{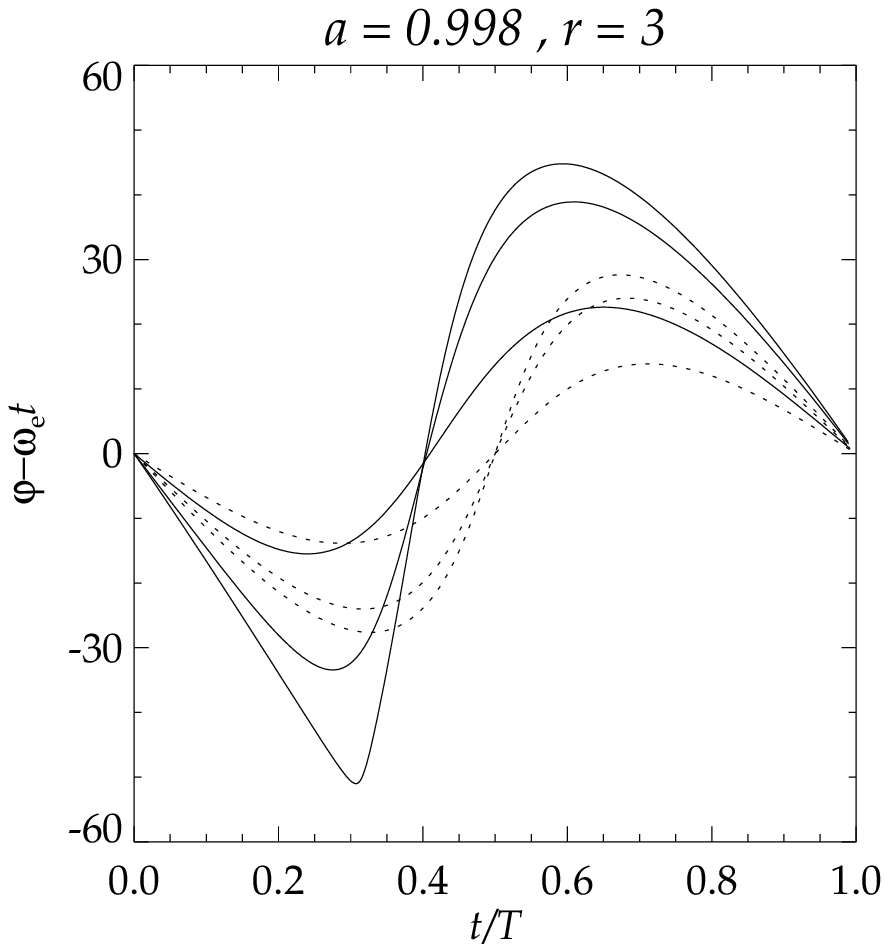}
  \hfill
  \includegraphics[height=4.7cm]{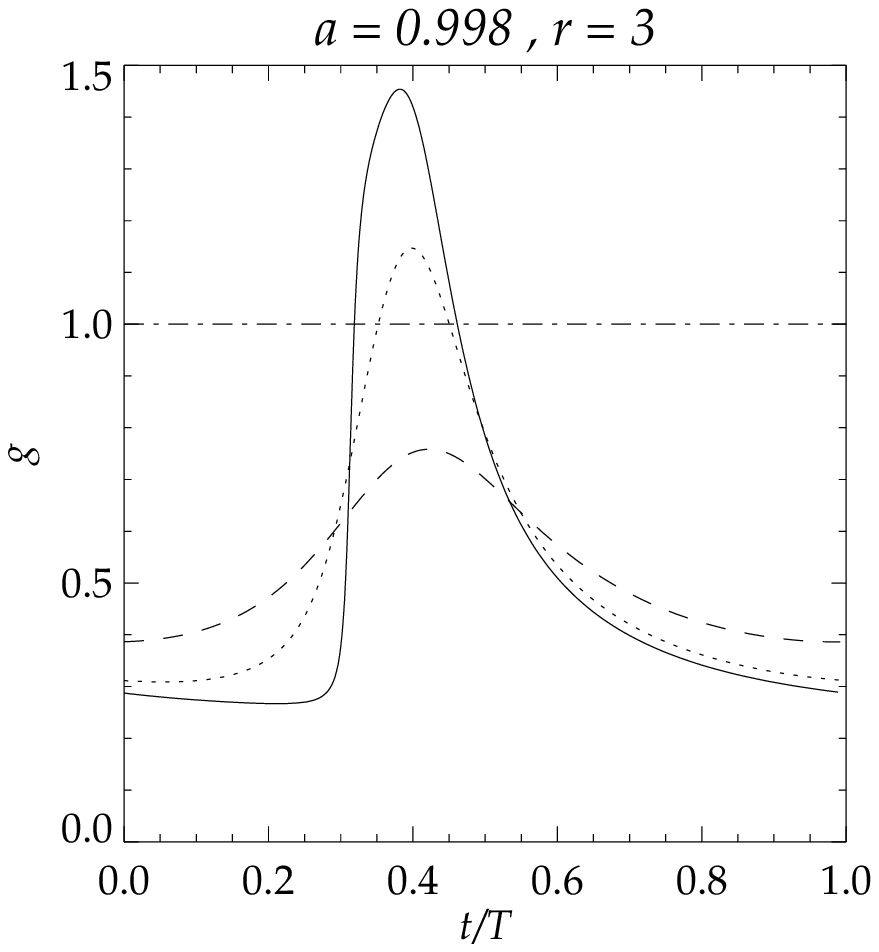}
  \hfill
  \includegraphics[height=4.7cm]{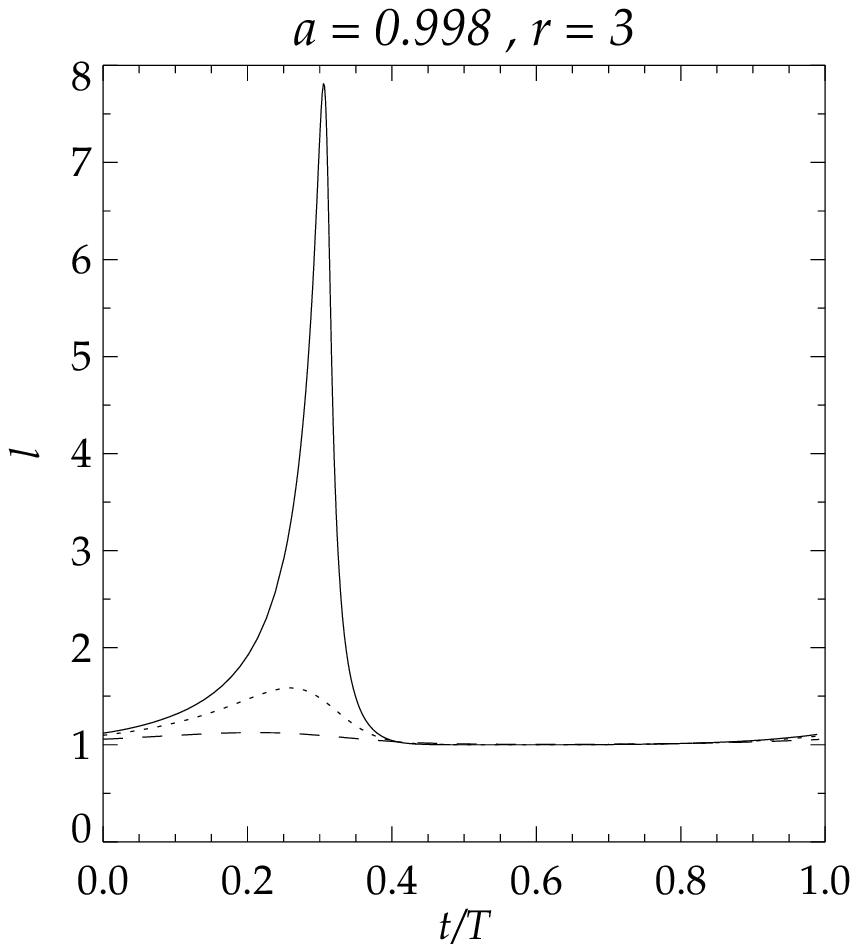}
  \hfill
  \includegraphics[height=4.7cm]{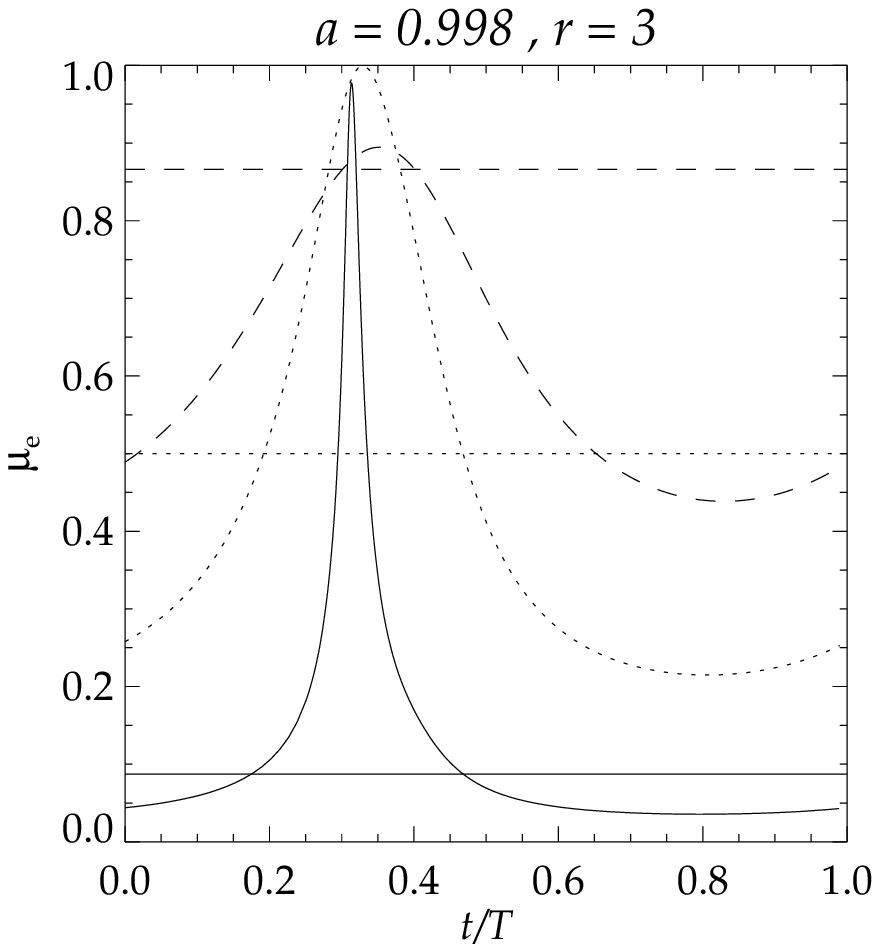}
  \vspace*{-2mm}
  \caption{Several important functions changing within one orbital timescale
   $T$ for the Schwarzschild (top panels) and the Kerr
   (bottom panels) black hole. The radius of the orbit is
   at $7\,GM/c^2$ for the Schwarzschild black hole and $3\,GM/c^2$ for the Kerr
   black hole. The initial
   time corresponds to the detection of the first photon. The flare and
   the spot are initially moving from the observer ($\varphi=0^\circ$).
   The dashed, dotted and solid
   lines correspond to the inclination of the observer 30$^\circ$,
   60$^\circ$ and 85$^\circ$.
   {\bf Left:} The lag (gain) angle $-\omega_{\rm e}\delta t$ due to the
   positive
   (negative) time delay $\delta t$ with which photons arrive to the observer.
   The solid and dotted lines correspond to the relativistic and
   classical cases, respectively. The lines for 30$^\circ$, 60$^\circ$
   and 85$^\circ$ are shown in the order of an increasing amplitude.
   {\bf Middle left:} The relativistic energy shift.
   {\bf Middle right:} The lensing.
   {\bf Right:} The cosine of the emission angle. The cosine of the observer's
   inclination angle is also shown by the straight lines.}
  \label{fig-tf1}
\end{figure*}
\begin{figure*}
\begin{center}
  \vspace*{-2mm}
  \includegraphics[height=4.7cm]{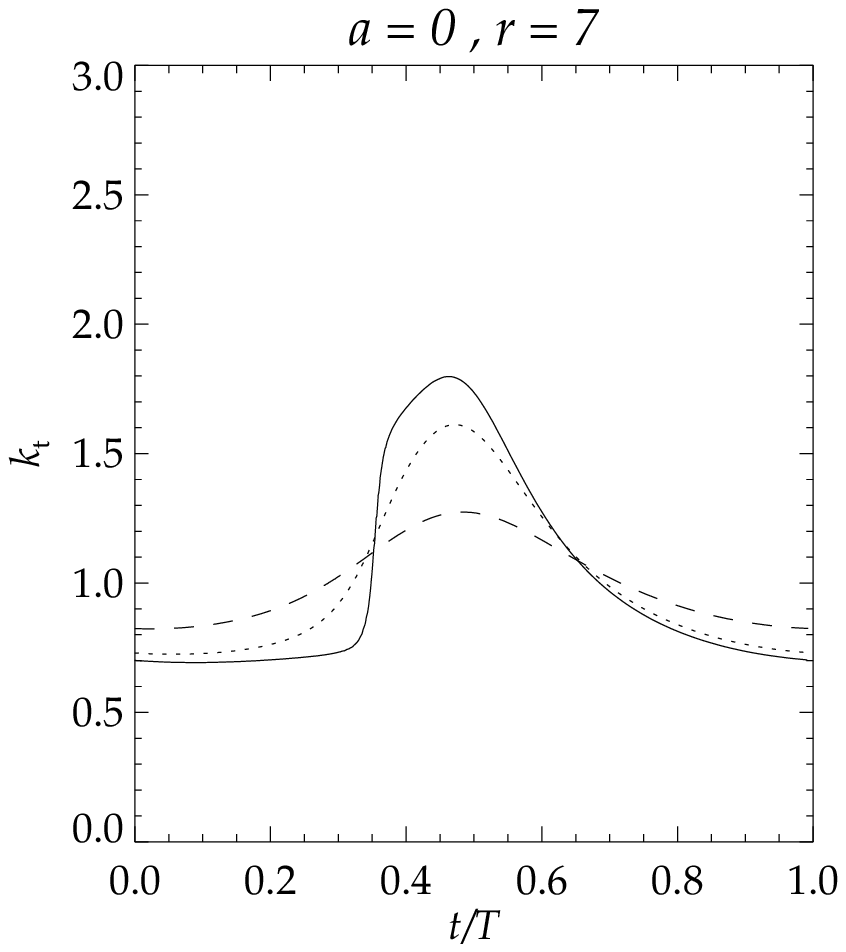}
  \hspace{5mm}
  \includegraphics[height=4.7cm]{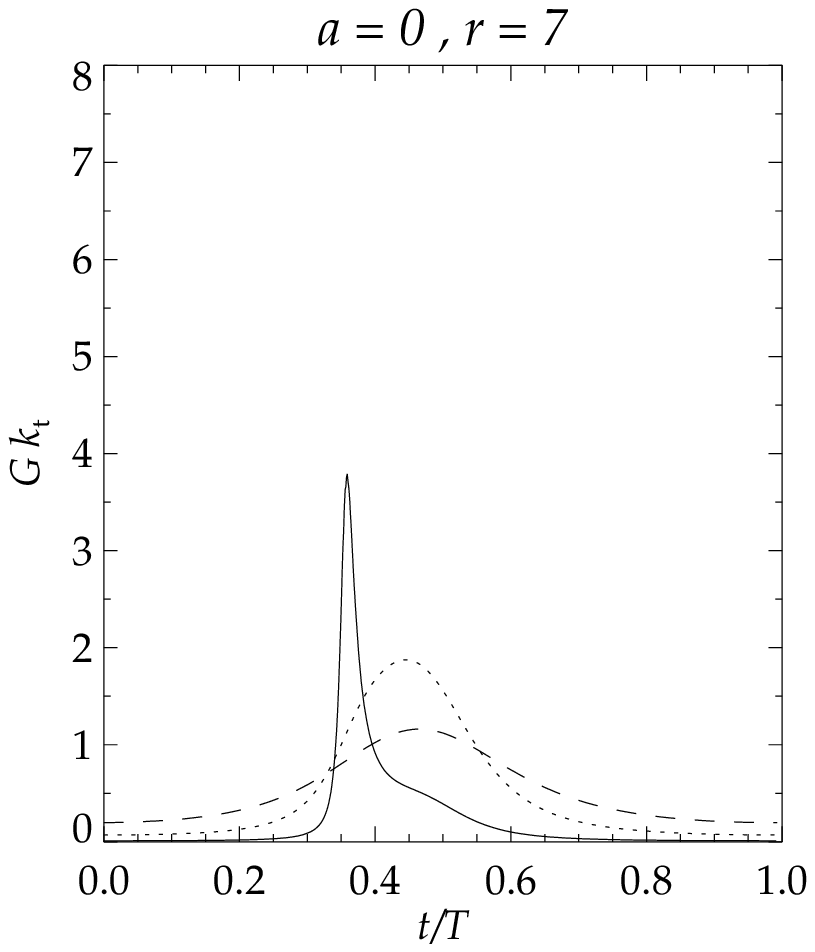}
  \hspace{5mm}
  \includegraphics[height=4.7cm]{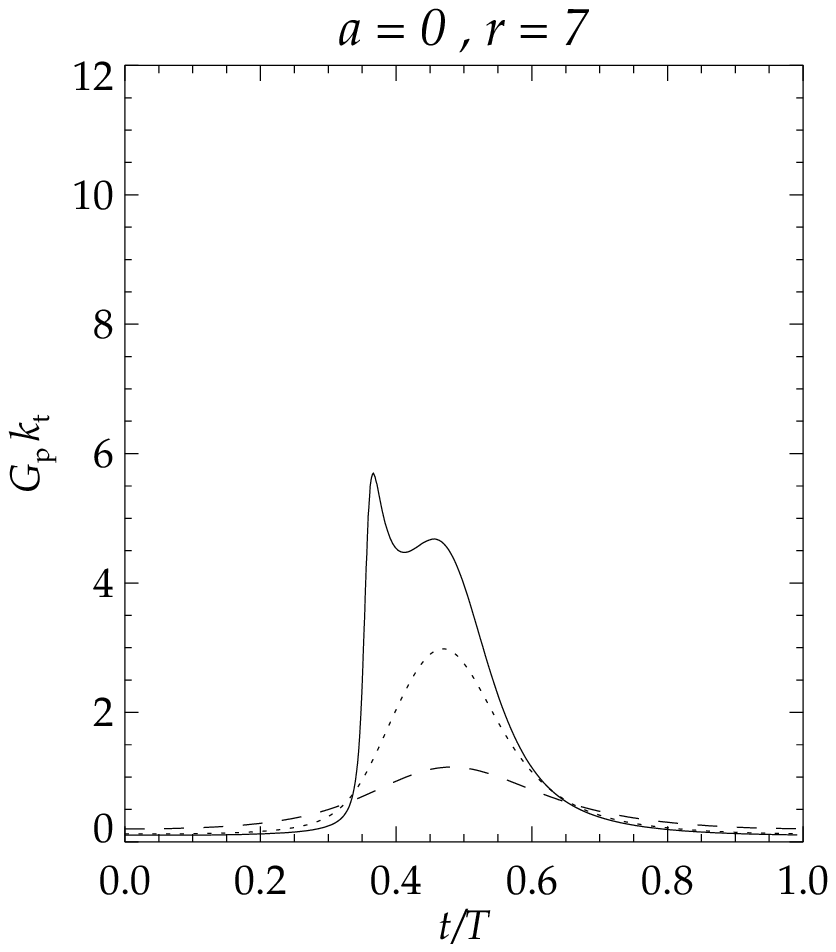}\\
  \includegraphics[height=4.7cm]{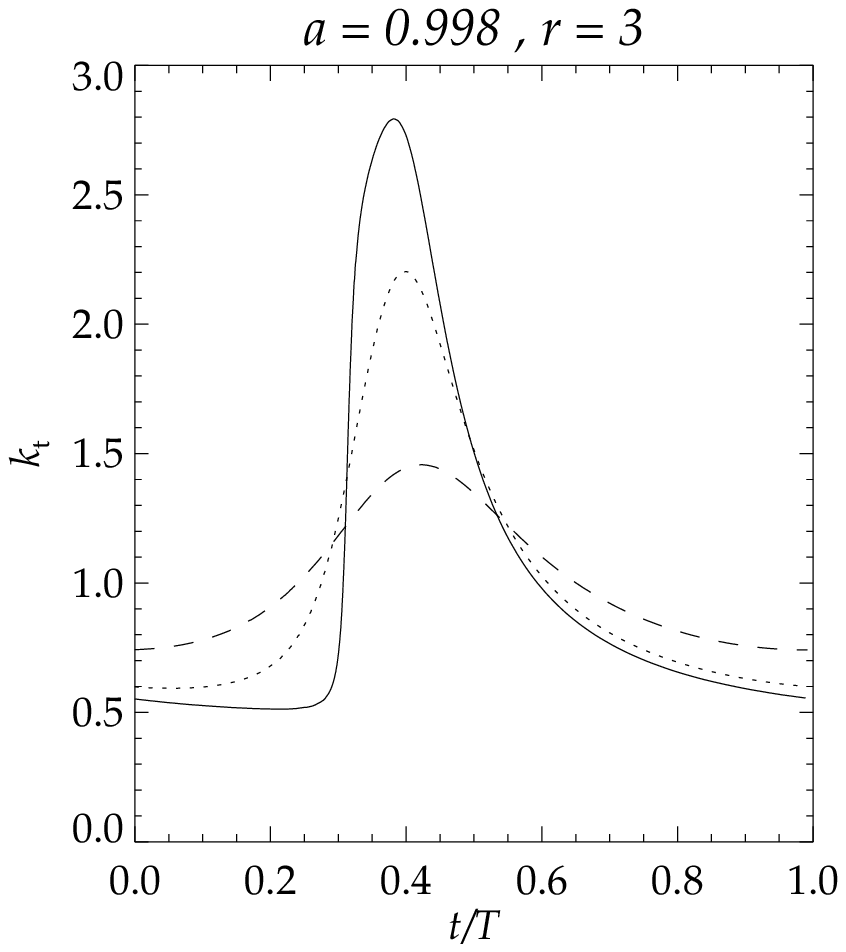}
  \hspace{5mm}
  \includegraphics[height=4.7cm]{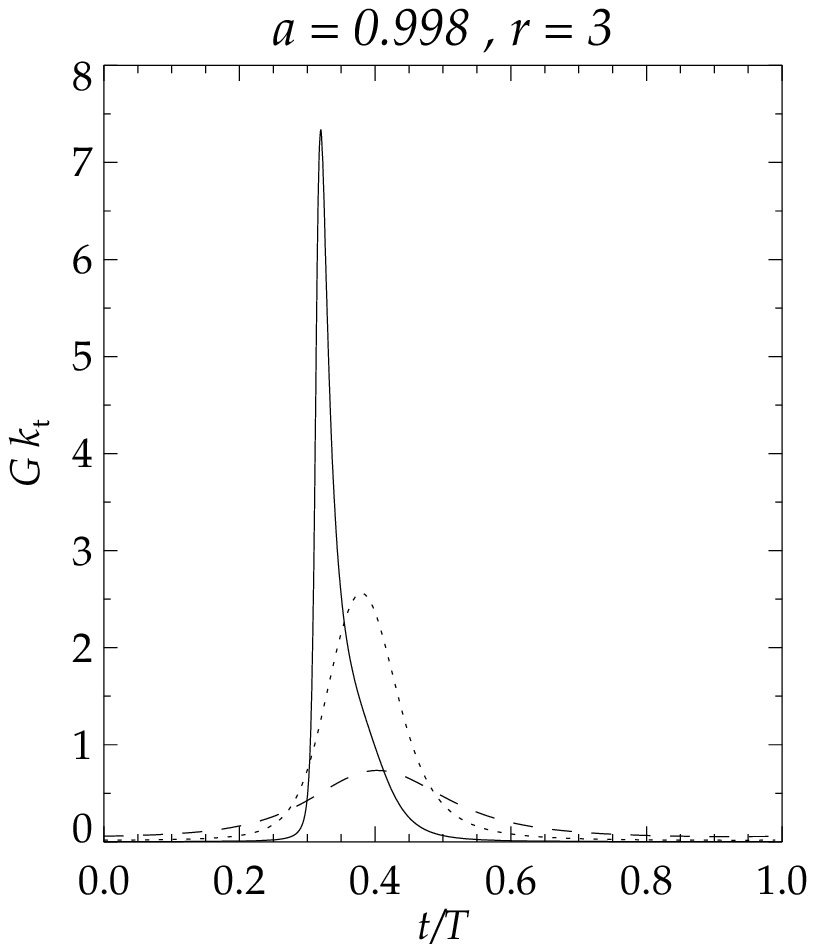}
  \hspace{5mm}
  \includegraphics[height=4.7cm]{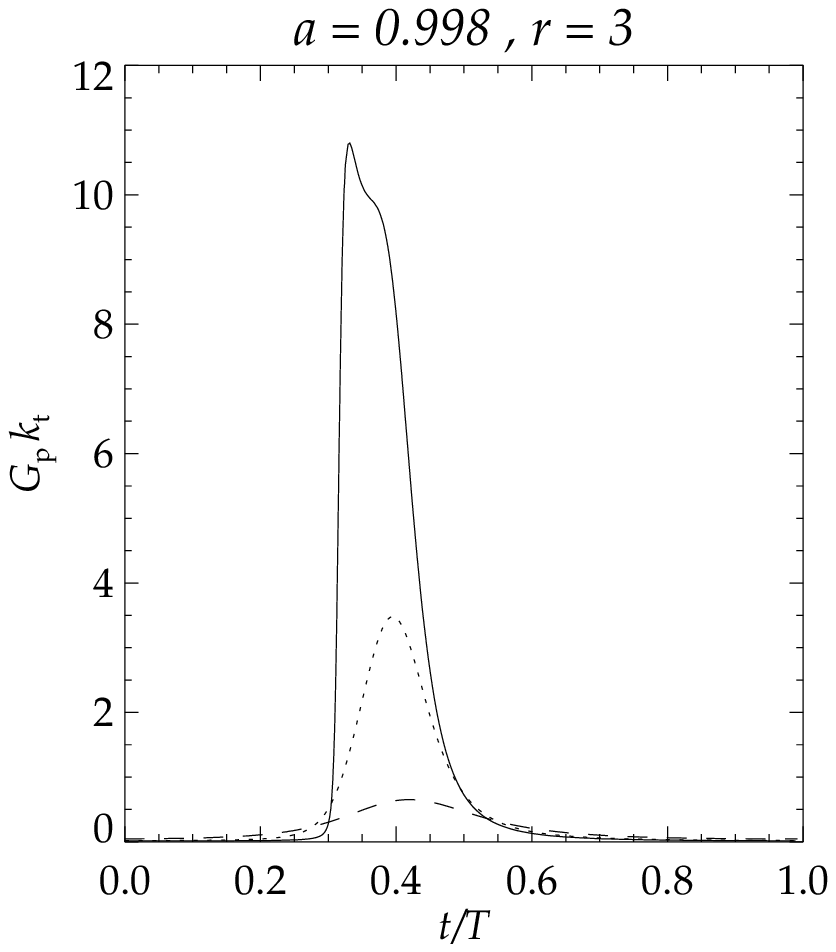}
  \vspace*{-2mm}
  \caption{{\bf Left:} Function $k_{\rm t}$ describing the time delay
   amplification.
  {\bf Middle:} The overall amplification of the local specific energy flux
  for an extended source of emission.
  {\bf Right:} The overall amplification of the local specific energy flux
  for a point source of emission.}
  \label{fig-tf2}
\end{center}
\end{figure*}

\begin{figure*}
\begin{center}
  \includegraphics[width=4.6cm]{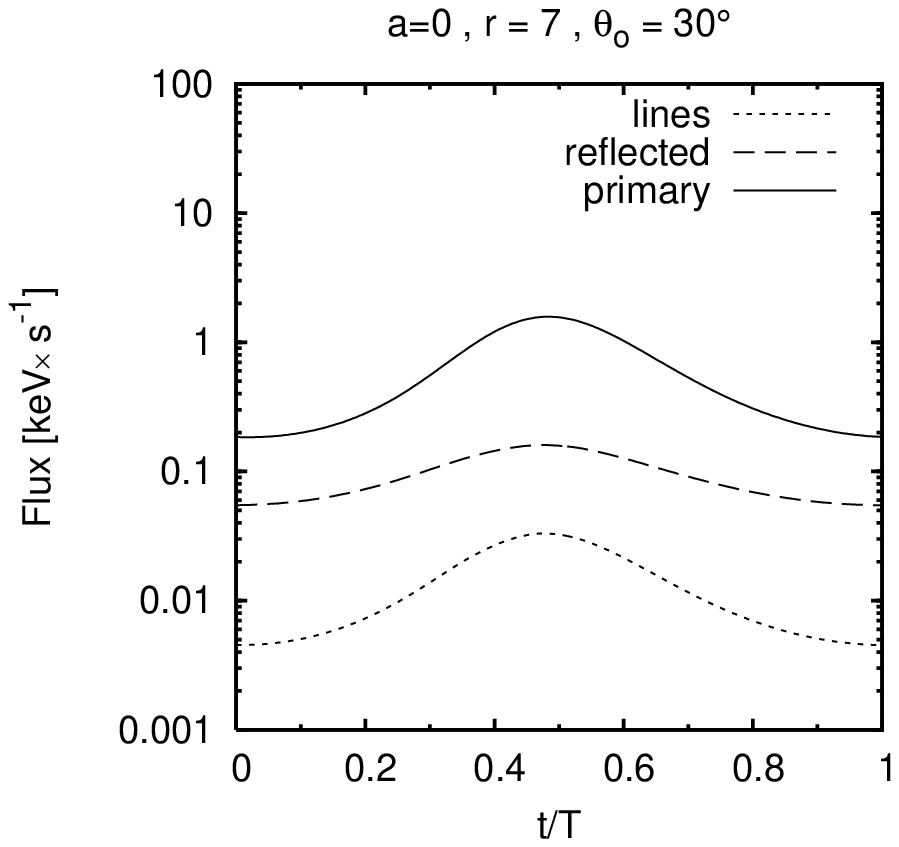}
  \hspace{5mm}
  \includegraphics[width=4.6cm]{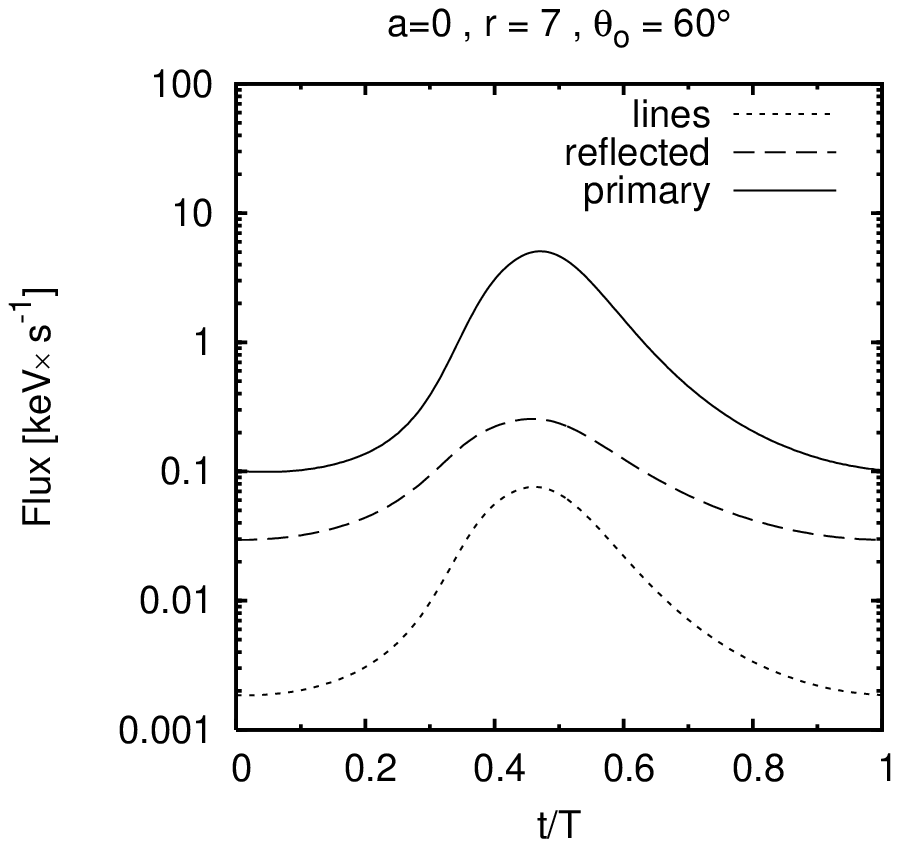}
  \hspace{5mm}
  \includegraphics[width=4.6cm]{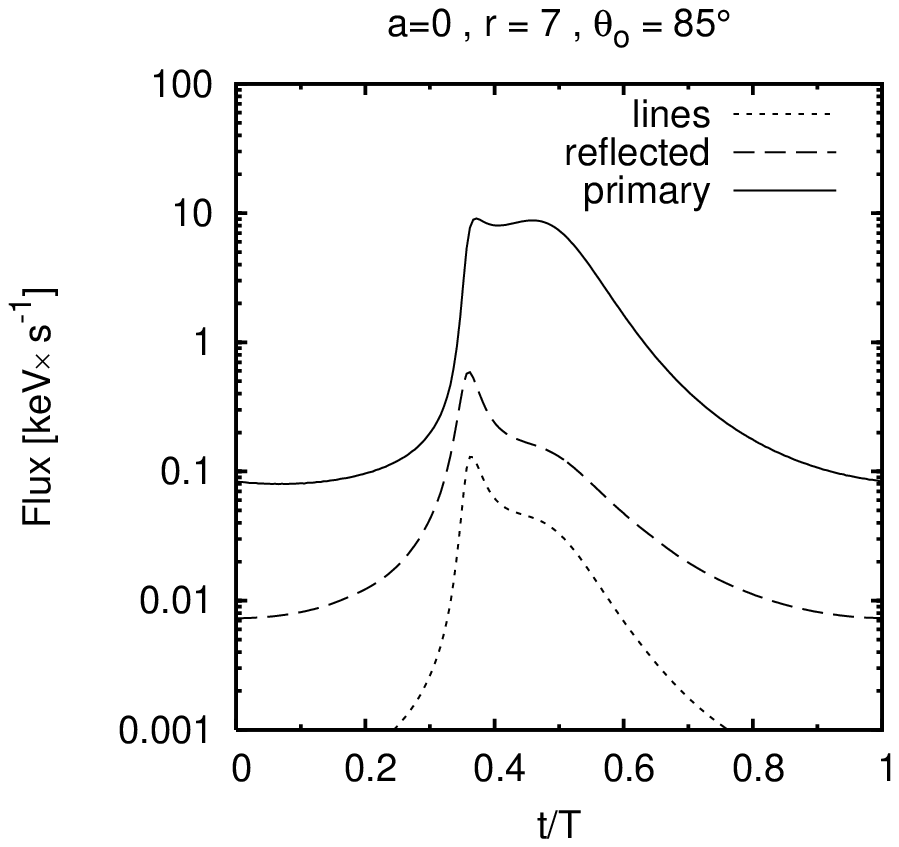}\\[4mm]
  \includegraphics[width=4.6cm]{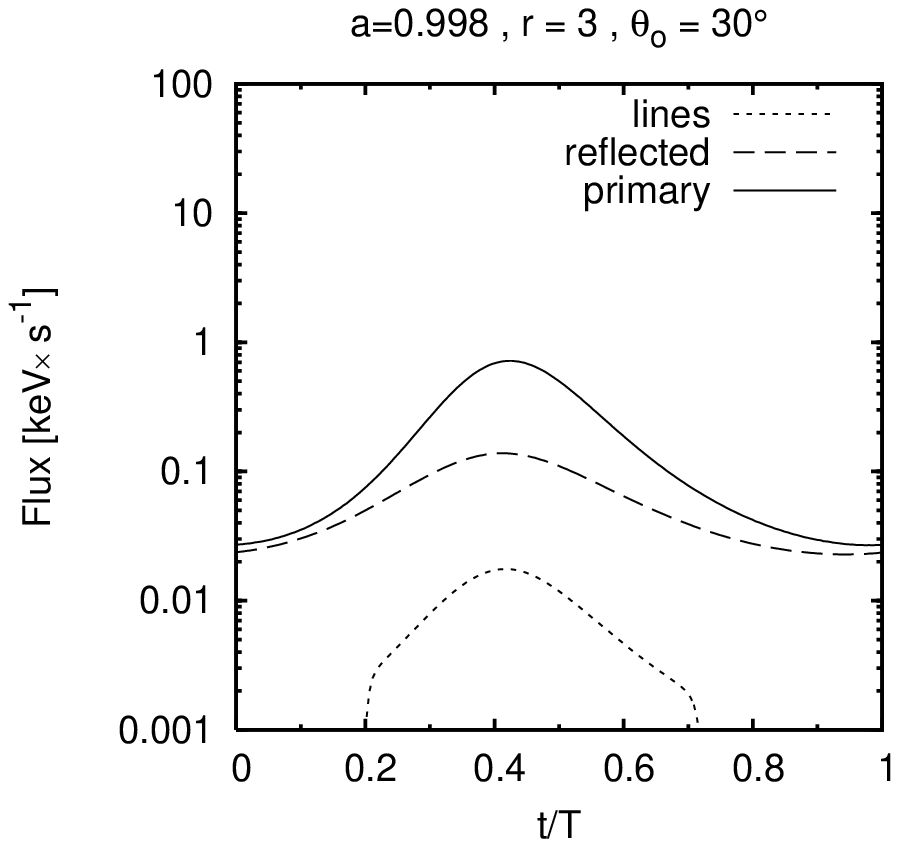}
  \hspace{5mm}
  \includegraphics[width=4.6cm]{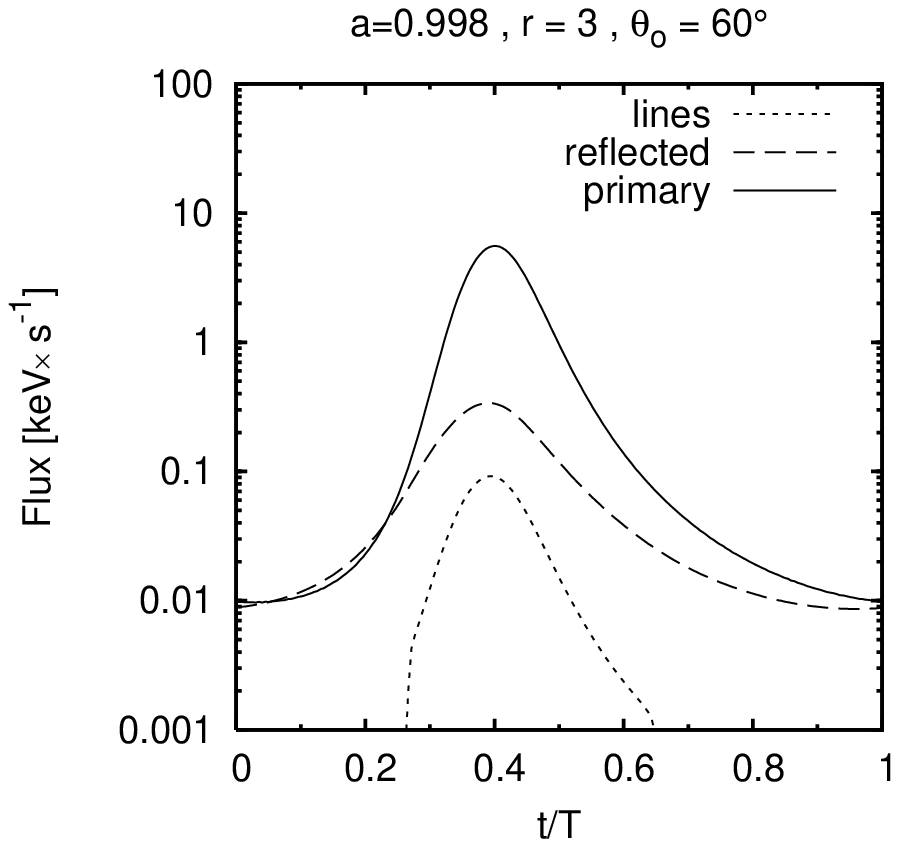}
  \hspace{5mm}
  \includegraphics[width=4.6cm]{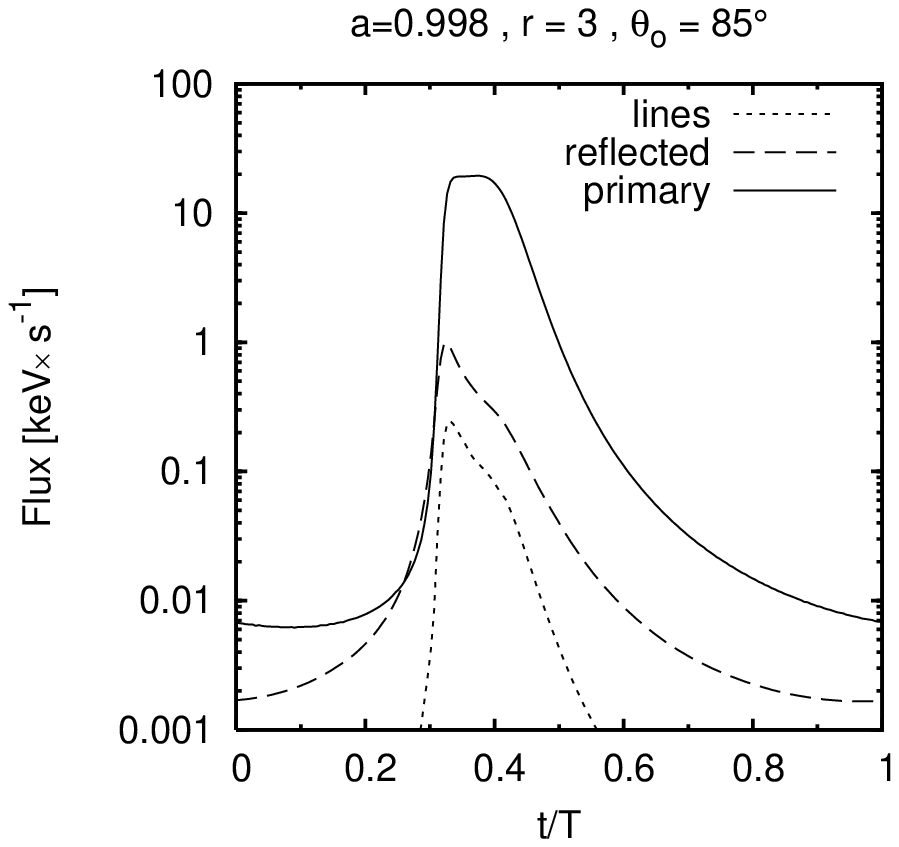}
  \caption{The light curves of the observed emission from the flare and the
  spot for the energy range 3--10~keV for the Schwarzschild (top) and the Kerr
  (bottom) black hole and observer's inclination angles 30$^\circ$,
  60$^\circ$ and 85$^\circ$ (from left to right). The primary emission, spot's
  continuum   emission and spot's emission in K$\alpha$ and K$\beta$ lines are
  denoted by solid, dashed and dotted graphs, respectively.}
  \label{fig-light_curves}
\end{center}
\end{figure*}
\begin{figure*}
\begin{center}
  \includegraphics[width=4.65cm]{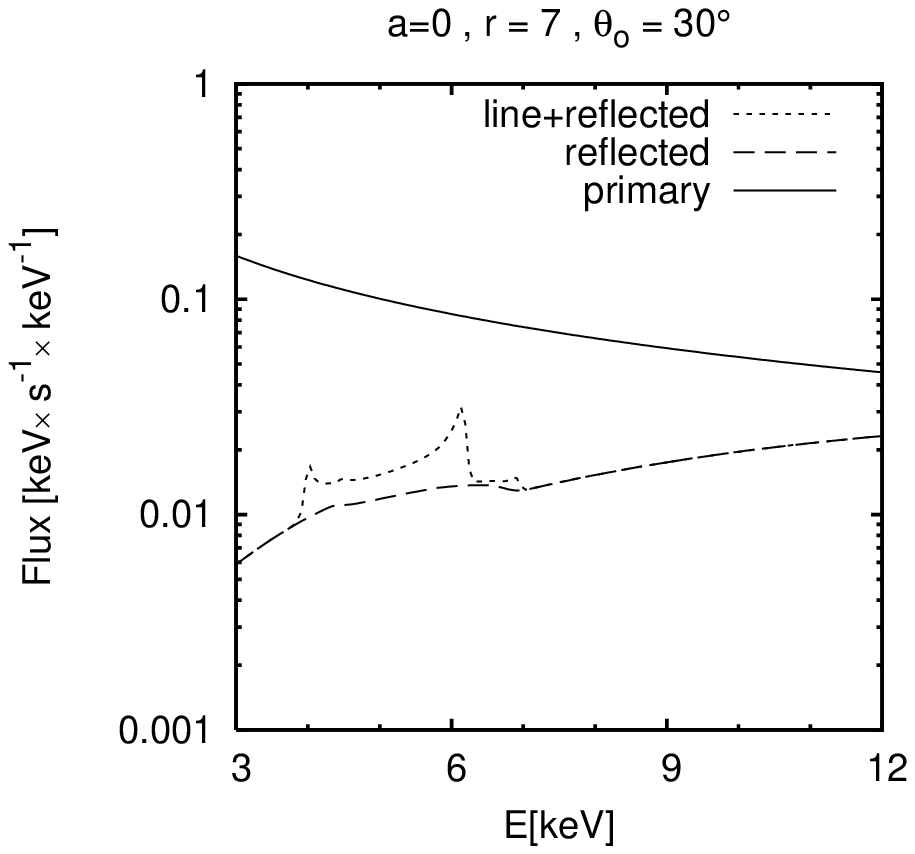}
  \hspace{5mm}
  \includegraphics[width=4.65cm]{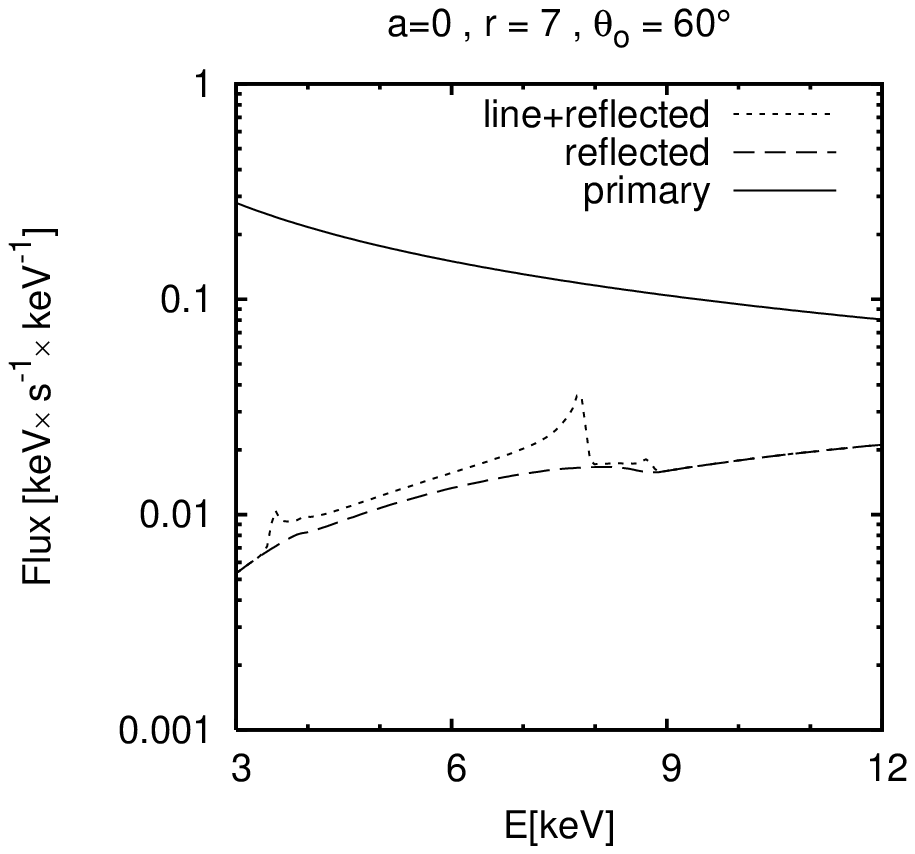}
  \hspace{5mm}
  \includegraphics[width=4.65cm]{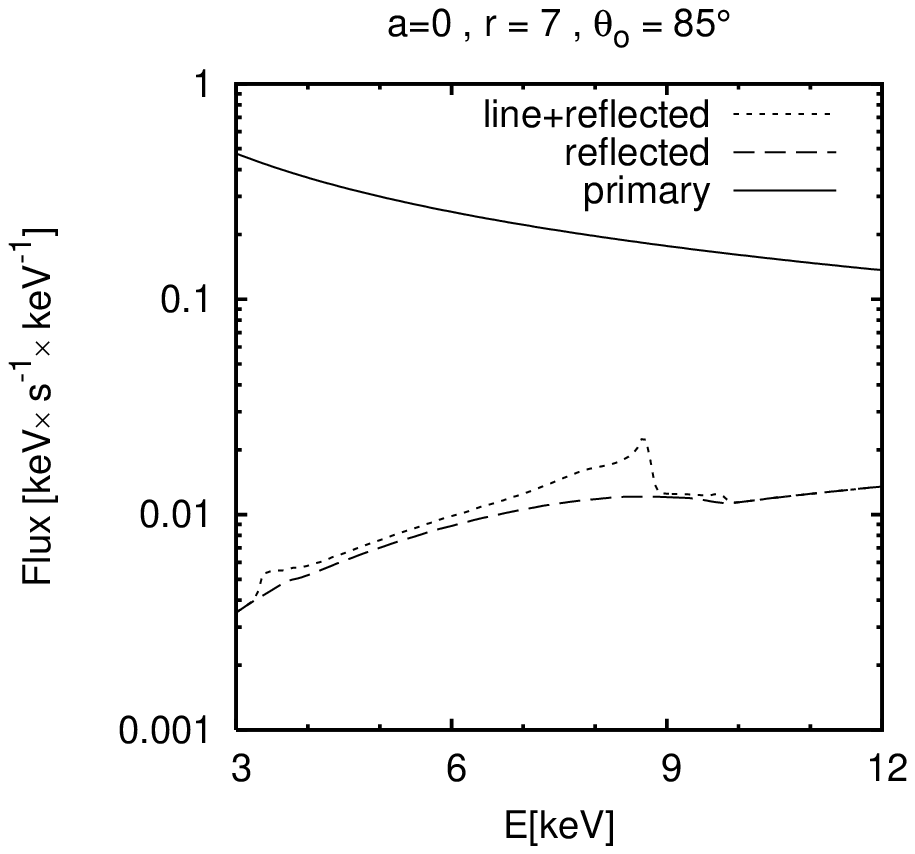}\\[4mm]
  \includegraphics[width=4.65cm]{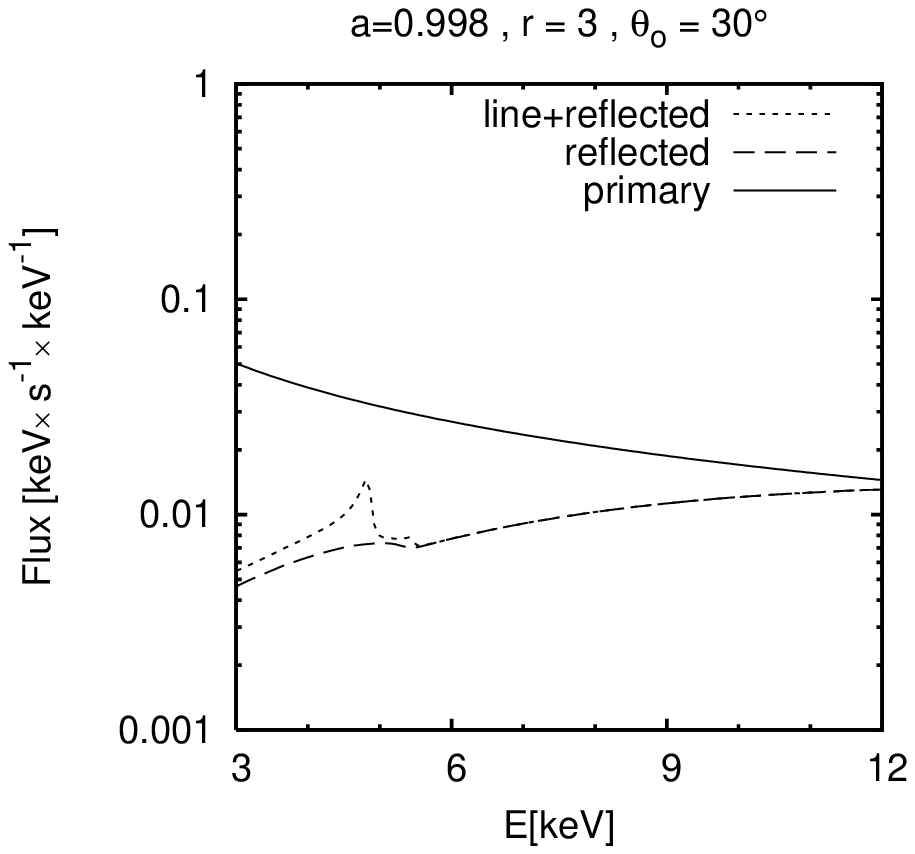}
  \hspace{5mm}
  \includegraphics[width=4.65cm]{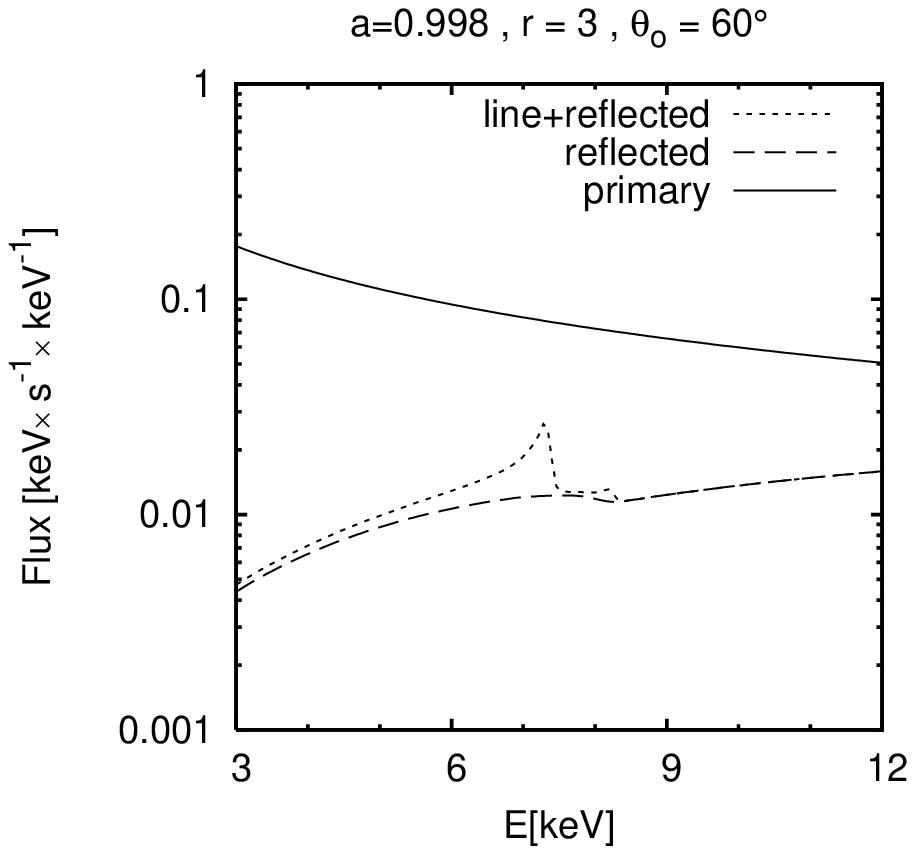}
  \hspace{5mm}
  \includegraphics[width=4.65cm]{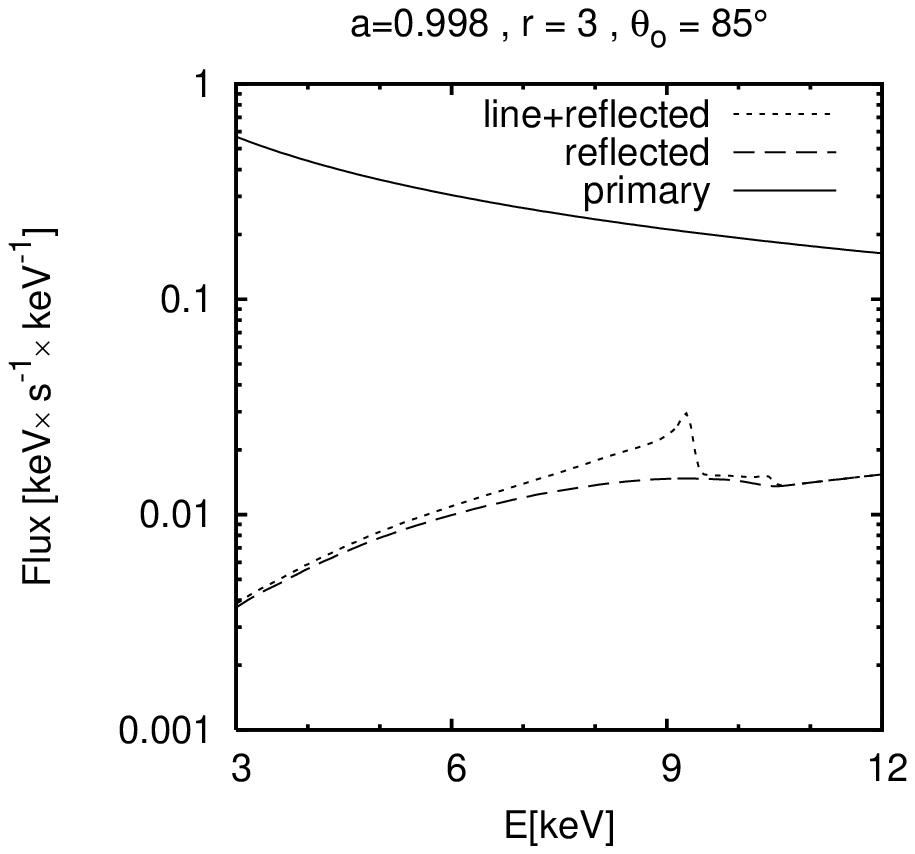}
  \caption{The observed spectra averaged over one orbit computed for the same
  set of parameters as in Fig.~\ref{fig-light_curves}. Here, the observed line
  flux is shown on top of the spot continuum emission.}
  \label{fig-spectrum}
\end{center}
\end{figure*}

\begin{figure*}
\begin{center}
  \vspace*{-3mm}
  \includegraphics[width=4.2cm]{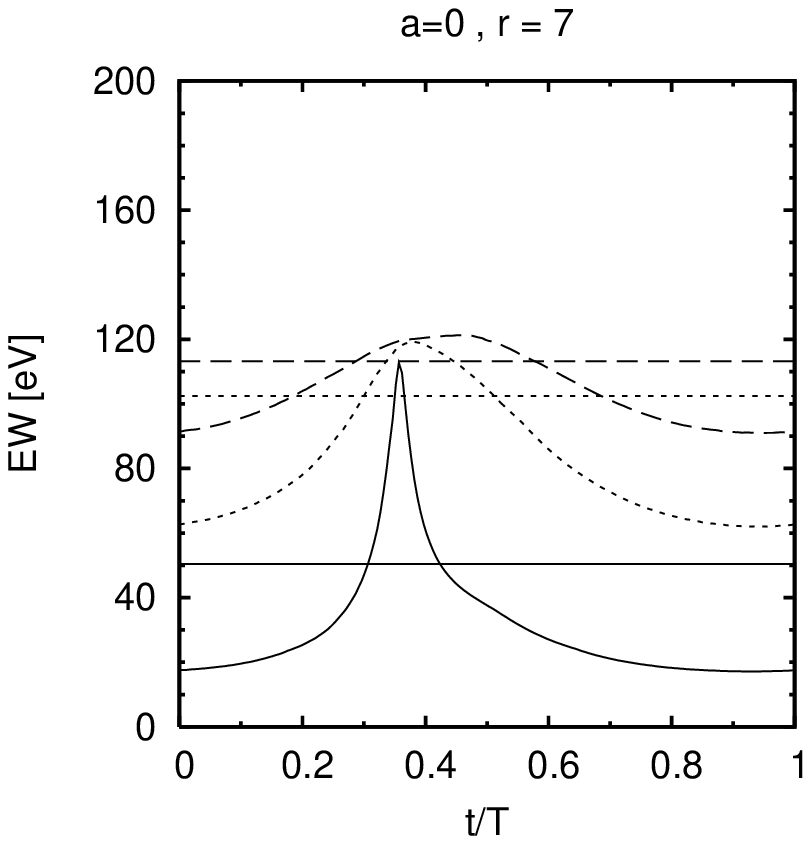}
  \includegraphics[width=4.22cm]{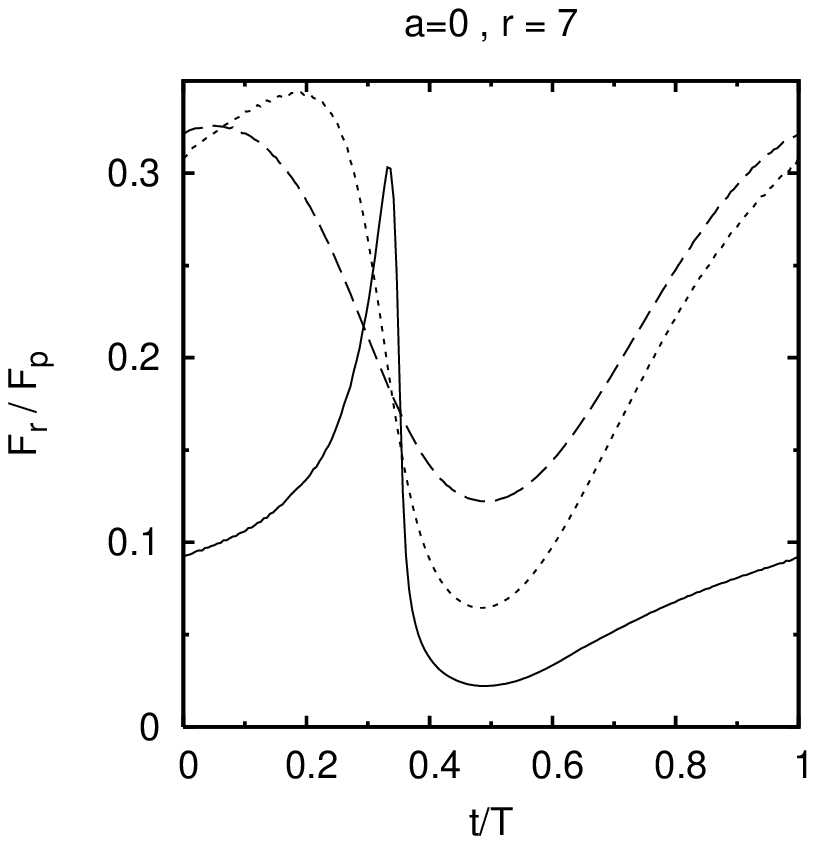}
  \includegraphics[width=4.25cm]{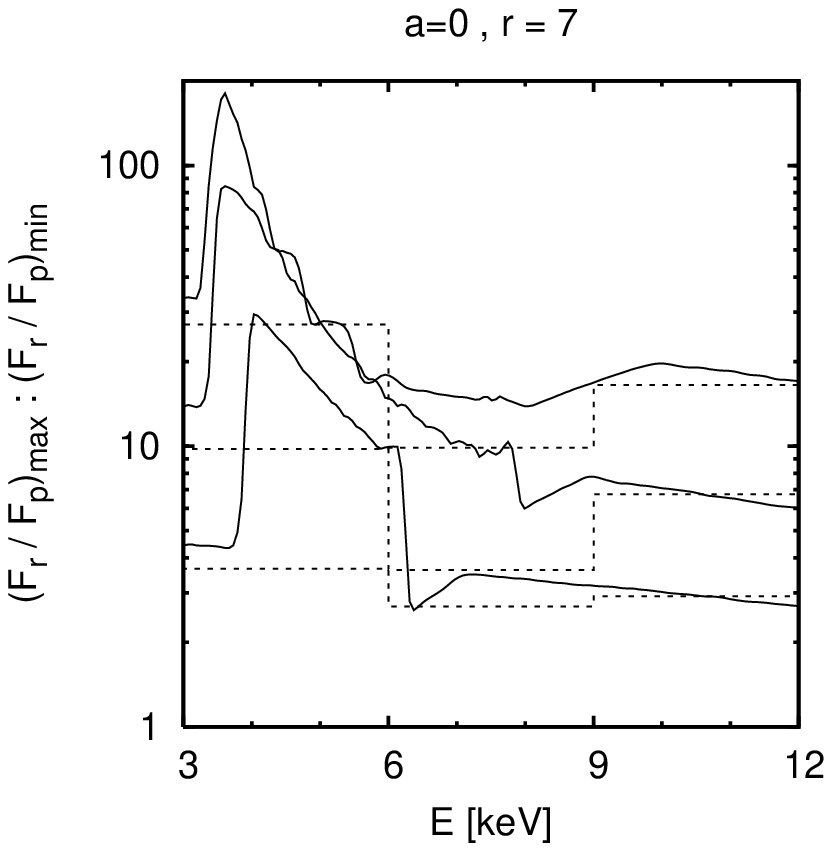}
  \includegraphics[width=4.27cm]{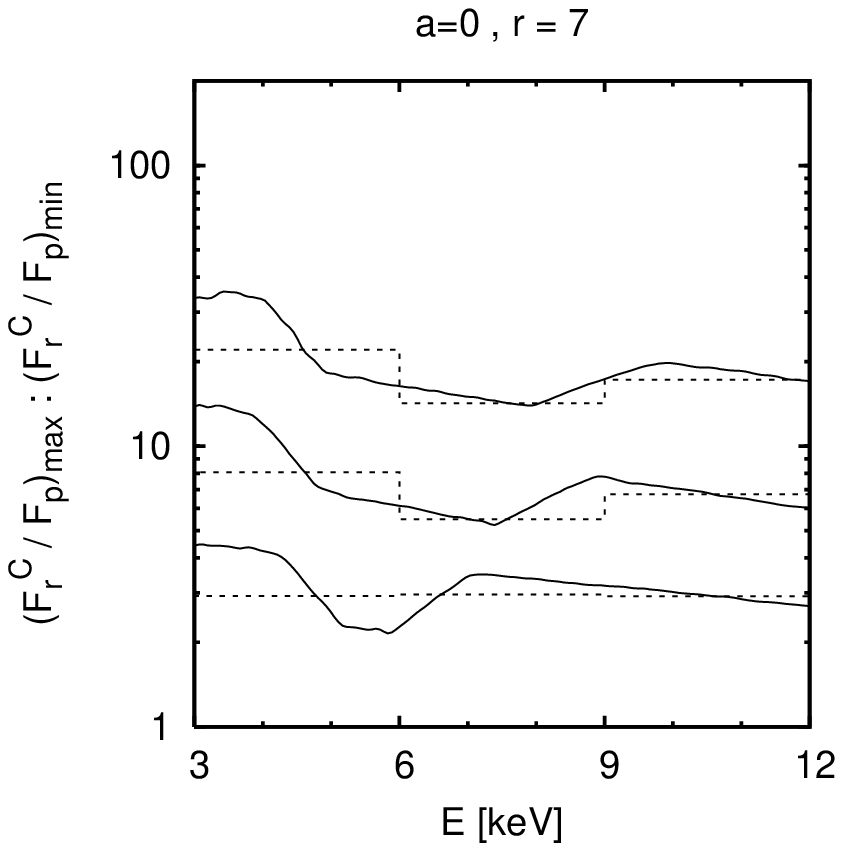}\\[4mm]
  \hspace*{0.07mm}
  \includegraphics[width=4.2cm]{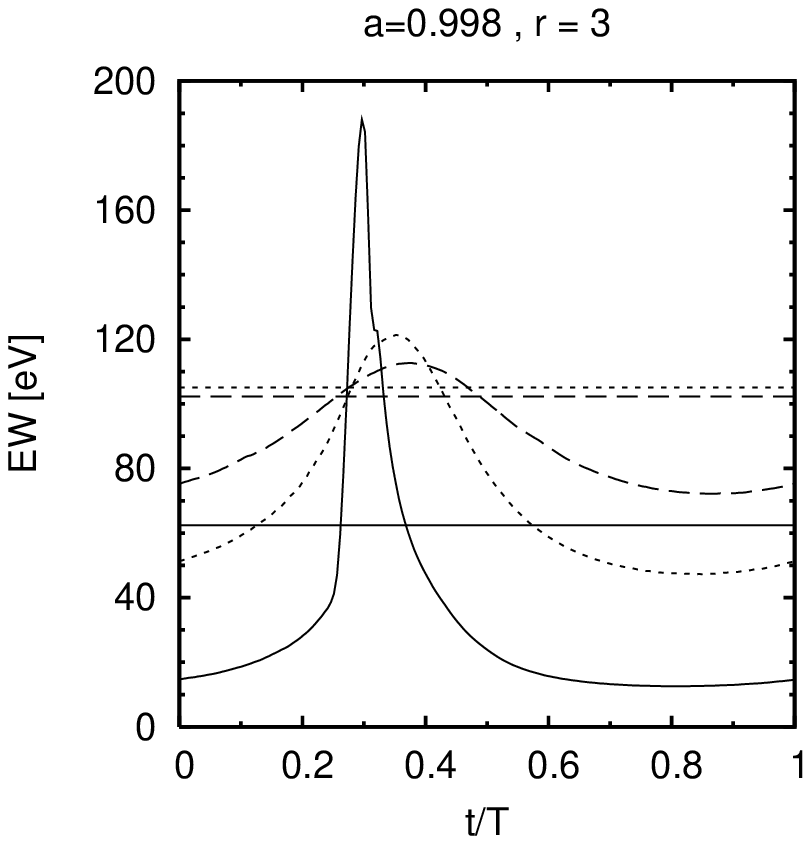}
  \includegraphics[width=4.22cm]{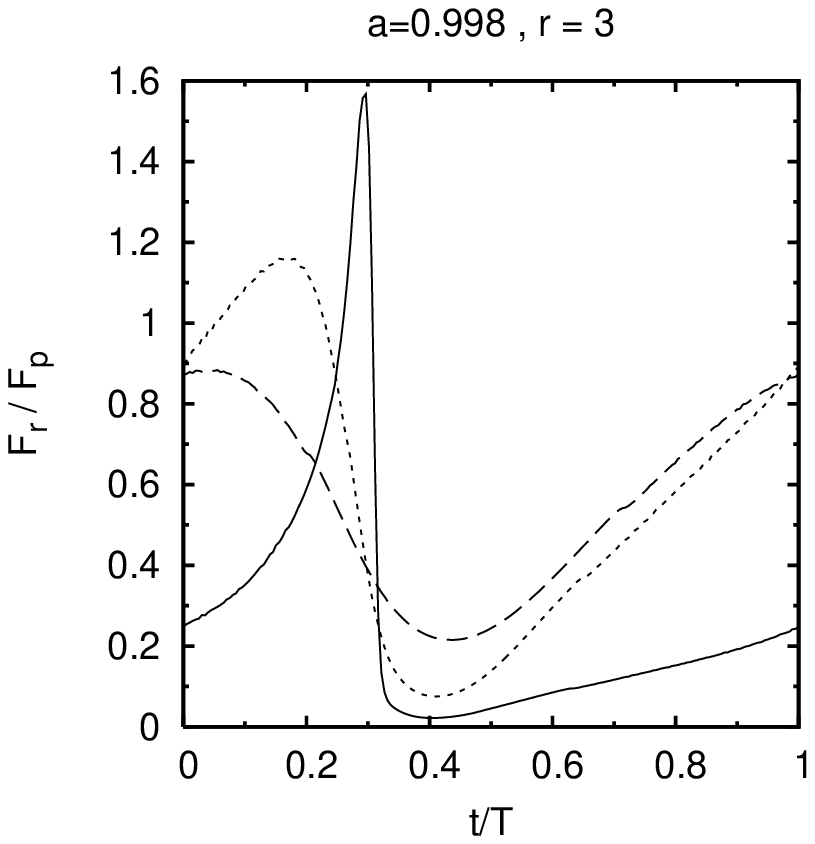}
  \hspace*{-2mm}
  \includegraphics[width=4.25cm]{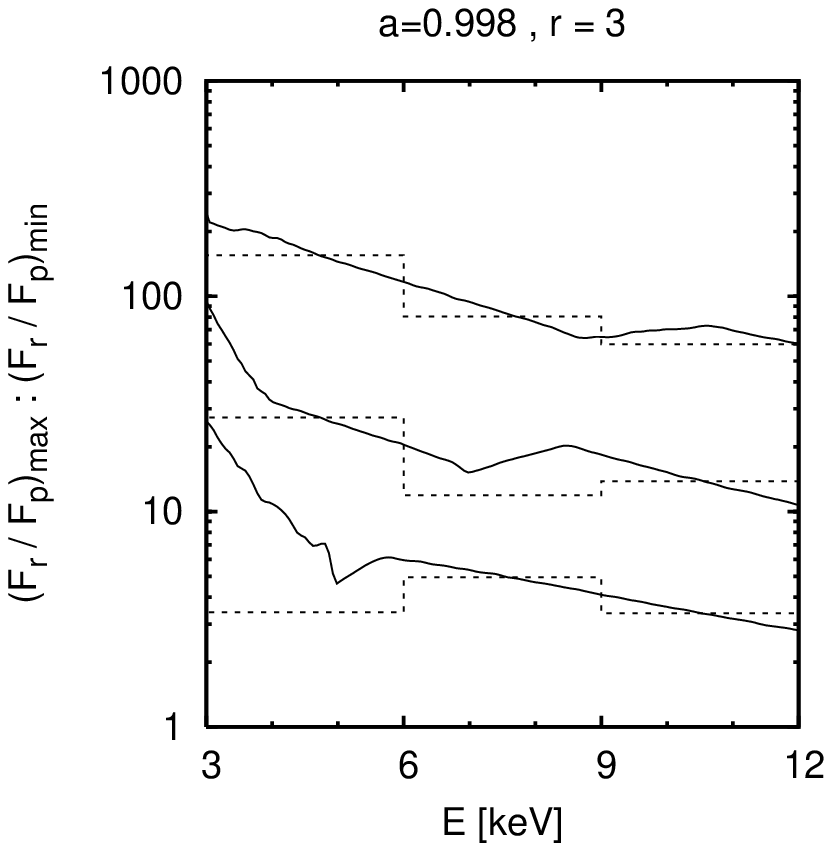}
  \hspace*{0.2mm}
  \includegraphics[width=4.27cm]{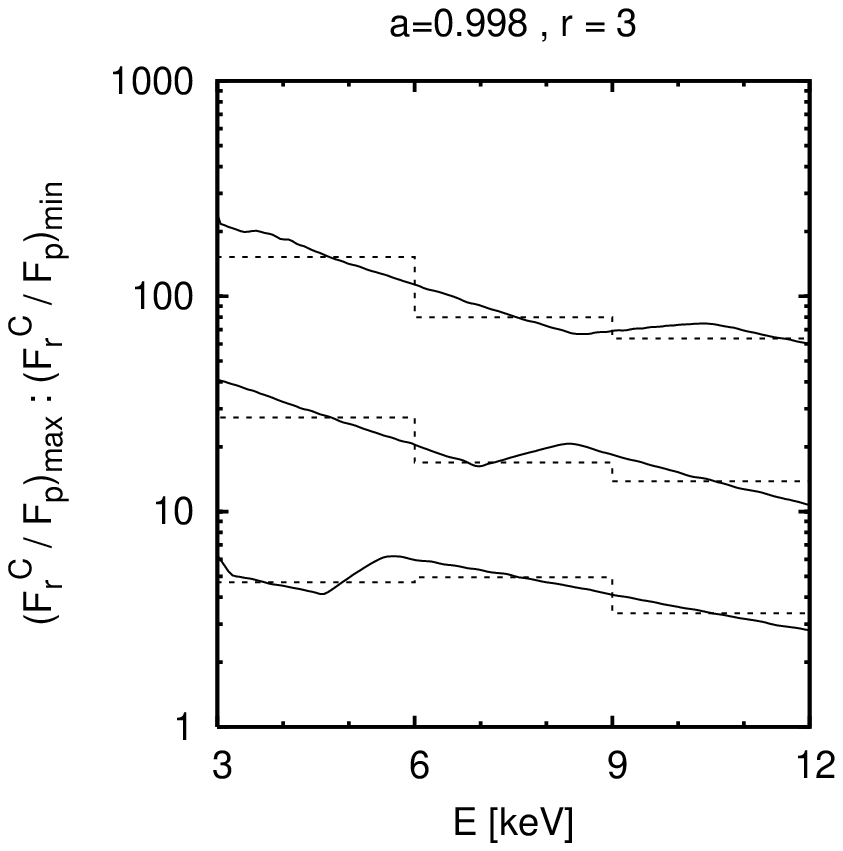}
  \vspace*{-2mm}
  \caption{{\bf Left:}
%  The observed equivalent width of K$\alpha$ line.
   The time variation of the observed equivalent width of the K$\alpha$ line
   is shown for three different observer's inclination angles --- 30$^\circ$
   (dashed), 60$^\circ$ (dotted) and 85$^\circ$ (solid). The EW integrated over
   the whole orbit is shown in horizontal lines.
  {\bf Middle left:} The ratio of the observed reflected emission to the
  observed primary emission. Both energy fluxes are integrated in the energy
  range 3--10~keV. The line styles are the same as in the left panel.
  {\bf Middle right:} The ratio of the maximum of the ratio of the observed
  reflected and primary emission $F_{\rm r}/F_{\rm p}$ to its minimum.
  The solid line represents the energy dependence of this ratio, the dotted
  line shows
  this ratio for fluxes integrated in the energy ranges 3--6~keV, 6--9~keV and
  9--12~keV. The inclination of the observer is $30^\circ$ (bottom lines),
  $60^\circ$ (middle lines) and $85^\circ$ (top lines).
  {\bf Right:} The same as in the middle right panel but without the flux
  originating in the Fe lines.
  The top (bottom) panels correspond to the Schwarzschild (Kerr) black hole.}
  \label{fig-ew}
\end{center}
\end{figure*}
\begin{figure*}
\begin{center}
  \vspace*{-6mm}
  \includegraphics[width=4.6cm]{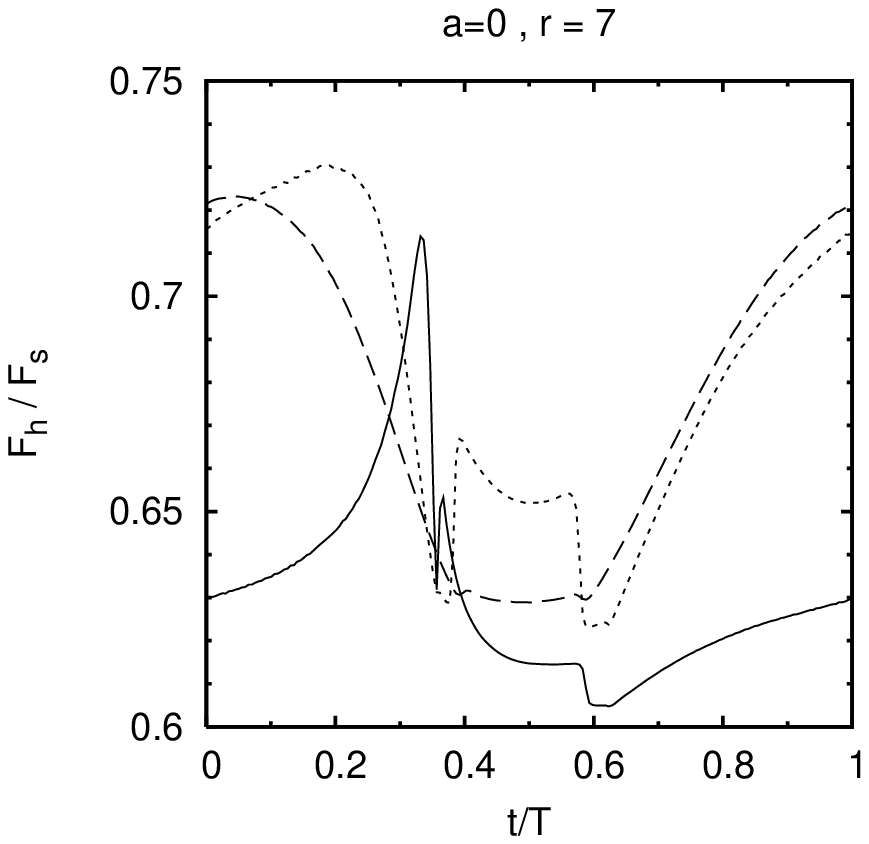}
  \hspace{5mm}
  \includegraphics[width=4.64cm]{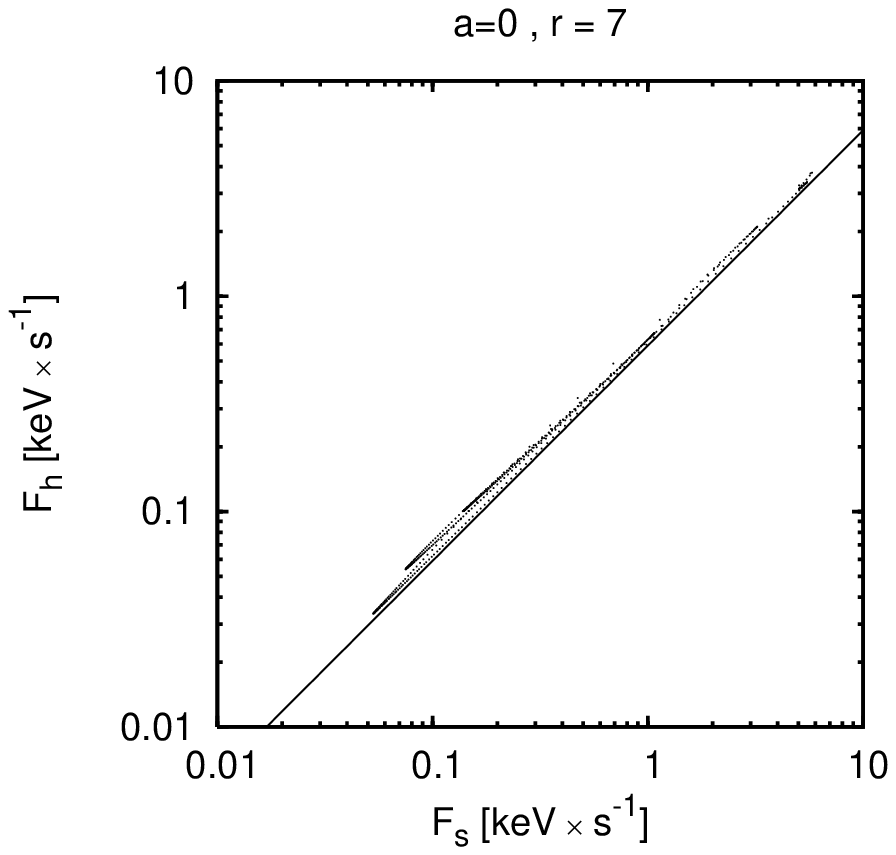}
  \hspace{5mm}
  \includegraphics[width=4.76cm]{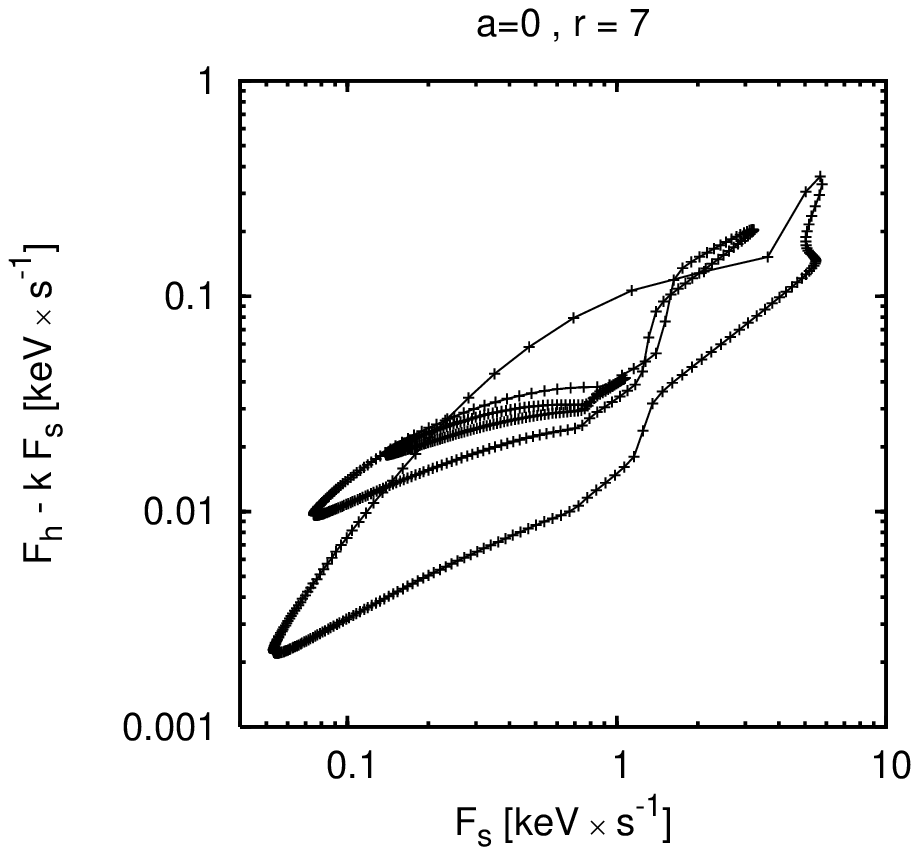}\\[4mm]
  \includegraphics[width=4.6cm]{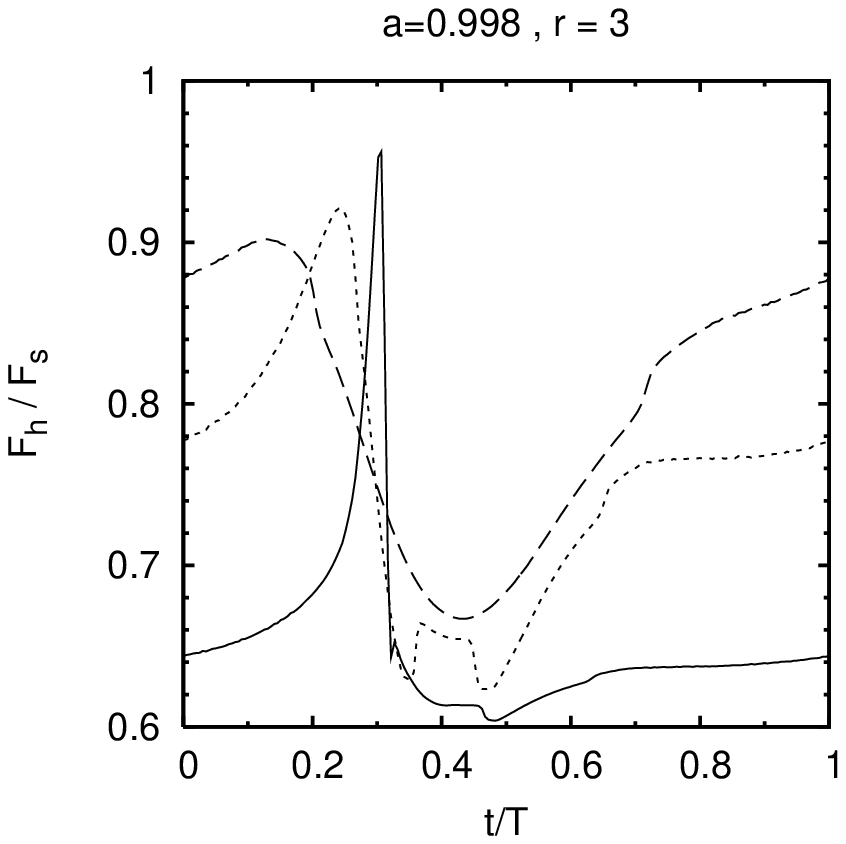}
  \hspace{2.5mm}
  \includegraphics[width=4.76cm]{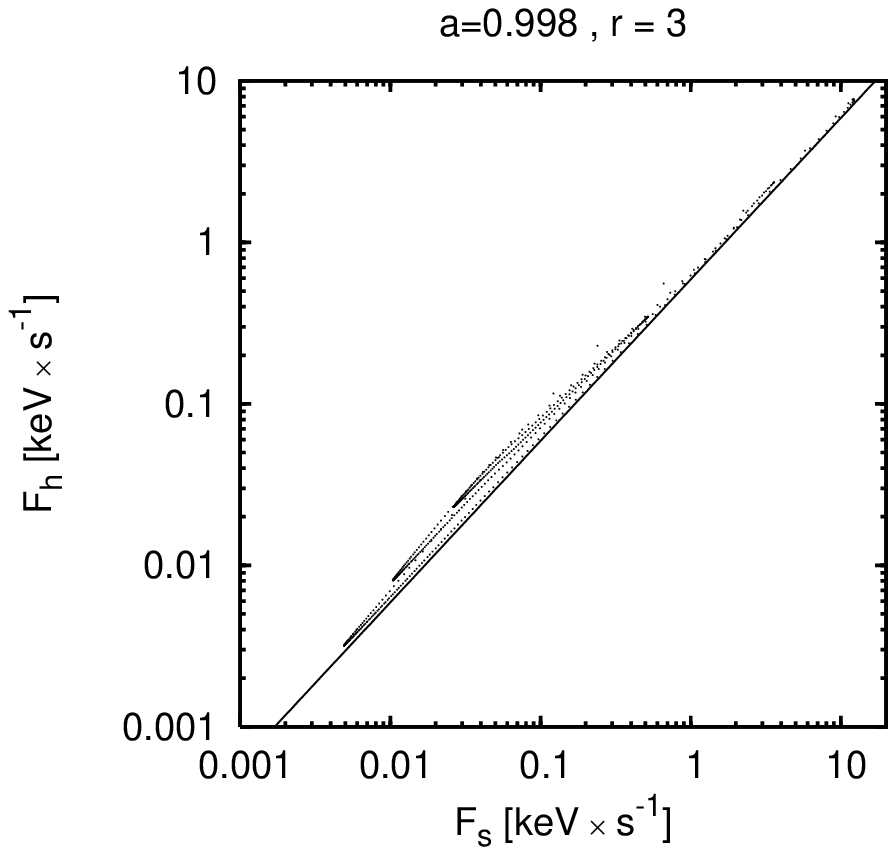}
  \hspace{4.5mm}
  \includegraphics[width=4.76cm]{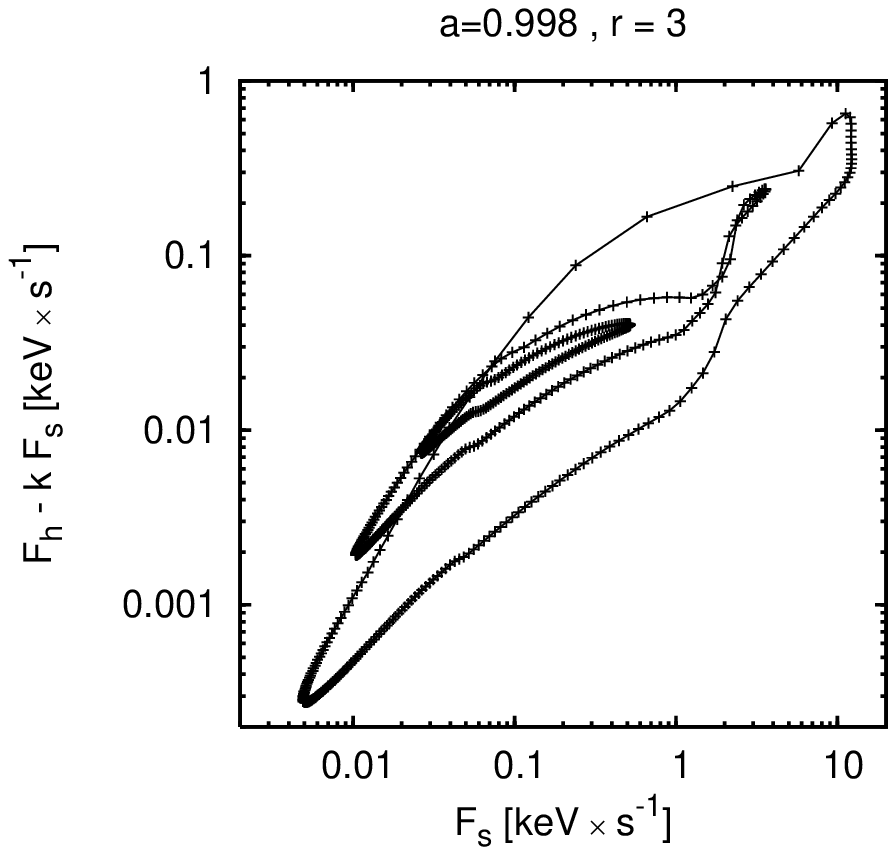}
  \vspace*{-2mm}
  \caption{{\bf Left:} The hardness ratio of the flux $F_{\rm h}$ in the
  energy range 6.5--10~keV to the flux $F_{\rm s}$ in the energy range
  3--6.5~keV.
  The flux in the Fe lines is also included. The dashed, dotted and solid
   lines correspond to the inclination of the observer being 30$^\circ$,
   60$^\circ$ and 85$^\circ$.
  {\bf Middle:} The hard flux $F_{\rm h}$ versus the soft flux $F_{\rm s}$.
  The solid line has a slope of $k=0.59$ (see text for more details).
  {\bf Right:} The same as in the middle panel but the line with the slope
  $k=0.59$ is subtracted from the hard flux. The loops are for different
  observer's inclinations (inner for $30^\circ$, middle for $60^\circ$ and
  outer for  $85^\circ$). The points are separated by the time $0.005\,T$.
  The top (bottom) panels correspond to the Schwarzschild (Kerr) black
  hole.}
  \label{fig-hardness_ratio}
\end{center}
\end{figure*}

\subsection{Effects of General Relativity}

The transfer function and delay amplification tell us how much the local
flux from the flare and the spot is amplified when observed at infinity.
Both of these functions depend on the location on the disc where the
observed photons are emitted. Therefore one needs to know where the spot
was at the time when a particular observed photon was emitted, in other
words one must know the dependence $\varphi(\varphi_{\rm e},t)$ given by
eq.~(\ref{phi}). If the spot is locally moving with a constant velocity,
the observer at infinity sees that the spot changes its
velocity, due to its motion and different time delays with which the emitted
photons are observed. Thus the spot seems to either fall behind or overrun
the position it should have if all photons arrive with the same time delay
(see the left panel in Fig.~\ref{fig-tf1} for $\varphi_{\rm e}=0$).
This is true even in the classical case.
% because the spot
%is moving towards or away from the observer thus
%diminishing or increasing the delay time of photons.
The differences between the classical and relativistic cases
can be very large especially for the closer orbits of the spot, larger
observer's inclination angle and faster rotating black hole.

The transfer function consists of several components --- the energy shift $g$,
the lensing $l$ and for an extended source also the emission angle
$\mu_{\rm e}$.
The contribution of each of these components to the transfer function for our
set of parameters can be seen in Fig.~\ref{fig-tf1} (taken for the position
of the flare or centre of the spot).

We show the magnitude of General Relativity effects for the Schwarzschild
($a=0\,GM/c^3$) black hole and the extremally spinning Kerr\footnote{We
use the value $a=0.998\,GM/c^3$ because it is
usually accepted as astrophysically the most extremal case of a rotating black
hole. The exact value for the extremal spin of the astrophysical black hole is
rather model-dependent although close to $0.998\,GM/c^3$ for a standard disc
(see Thorne \citeyear{tho74}).
%We found out that the results for $0.995\,GM/c^3<a<1\,GM/c^3$ differ
%less than by 2\%.
} ($a=0.998\,GM/c^3$) black hole, assuming different
inclinations of the observer and setting the primary flare at
a small height $h$ above the disc (typically $h=0.015\,GM/c^2$, resulting
in the spot radius $0.86\,GM/c^2$). The power-law photon index of the primary
radiation is $\Gamma=1.9$ and the spot is illuminated by photons
emitted within a downwards directed cone (with the half-opening angle
$89^\circ$).

The energy shift, $g$-factor, can either amplify or diminish the
observed flux. The latter is the case for the whole orbit for the inclination
$\theta_{\rm o}=30^\circ$ which is due to the fact that the spot orbits very
close to the black hole and gravitational redshift prevails over the Doppler
shift.

The lensing effect has the largest amplification
contribution for the inclination $85^\circ$ whereas we can neglect it
for the lowest inclination of $30^\circ$. In the middle case
($\theta_{\rm o}=60^\circ$) its effect is comparable to other amplification
parameters.

The emission angle is important for the extended source not only as part of
the transfer function but also as a variable that the reflected component
$f_{\rm r}$ of the local flux depends on heavily, as can be seen from the
left panel in the
Fig.~\ref{fig-ratio_loc} --- the local flux changes by one order of magnitude
for the full range of angles. The emission angle can change dramatically for
higher inclinations during the whole orbit, and for $\theta_{\rm o}=85^\circ$
in the Kerr case it acquires almost all possible values.

The delay amplification plays an important role in modifying the local flux
as can be seen from the left panels in Fig.~\ref{fig-tf2}. As described
earlier, according to the observer, the spot spends more time in certain parts
of the orbit than in the others and, as a consequence, less photons per unit
observers time is detected.
%This also means that by approximating emission from
%moving spot by steady emission from a patch of the disc we introduce large
%inaccuracies (by omitting the $k_{\rm t}$ factor).
The $k_{\rm t}$ factor has a larger maximum in the Kerr case. It is mainly due
to the fact that the spot orbits with a larger velocity being closer to
the black hole.

From the middle and right panels in Fig.~\ref{fig-tf2} one can see what is the
overall observed amplification of the local flux from the centre of the spot
(extended source) and from the flare (point source). Because the overall
amplification of the flare's flux has one $g$-factor more, the flux from the
flare is more amplified where $g>1$ and less in the other case. For larger
inclinations the overall amplification has a larger maximum in the Kerr case,
because of the larger lensing effect, Doppler shift and delay amplification,
whereas for lower inclinations it is lower because of the much smaller
gravitational shift.

\subsection{Spectral characteristics of the observed signal}

The observed light curves computed for the 3--10~keV energy range can be seen
in the Fig.~\ref{fig-light_curves}. The light curves are influenced mainly by
the overall amplification factor (transfer function and delay amplification)
and by the dependence of the local flux on the emission angle. The primary
emission dominates the observed flux as expected, meanwhile the reflected flux
in the Fe lines from the spot contributes less. There is an exception in
this behaviour, though, for some parts of the orbit in the Kerr case, when the
reflected flux from the spot exceeds the flux of the primary (see cases
$\theta_{\rm o}=60^\circ$, $85^\circ$).

Fig.~\ref{fig-spectrum} shows the mean spectra taken over the whole orbit.
The line is smeared when taken over the whole orbit.
As it is well known (Iwasawa et al.\ \citeyear{iwa96}) in the Schwarzschild
case the line stays above 3~keV, while in the Kerr case it can be shifted even
below this energy (as is the case for all inclinations for the spot orbit at
$3\,GM/c^2$). The iron edge is smeared in all studied cases and the dominance
of the primary emission is evident.

In order to quantify the properties of the observed spectra let us look at the
equivalent width, ratio of the observed reflected and primary components, and
the hardness ratio (Figs.~\ref{fig-ew} and \ref{fig-hardness_ratio}).

A closer look at the EW reveals that it does not much differ from its
local value except for the Kerr case with an observer
inclination of $85^\circ$. From eq.~(\ref{ew2}) it follows that the
observed EW should be equal to the local one multiplied by the
$g$-factor for a small spot, but this is not true in our case when we take
also the primary emission into account. This is due to the fact that
the relationship $Gk_{\rm t}\approx G^{\rm s}k_{\rm t}^{\rm s}$ does
generally not hold
true. Only in the reflection the factors $Gk_{\rm t}$ for the continuum
and the line components cancel each other, hence if we compute the EW for
the reflected emission only, then it does behave as the local one times $g$.

For the spot close to the black hole ($r=3\,GM/c^2$) the EW is changing
with respect to its mean value by 30\% even for a low inclination angle
$30^\circ$. For an almost edge-on disc it can vary by as much as 200\%.

The observed ratio of the reflected flux to the primary flux is amplified when
compared to the local one. Again, the approximation of a small spot cannot be
used. The amplification is the highest in the Kerr case with
$\theta_{\rm o}=85^\circ$ --- the ratio is increased by more than one order of
magnitude. Note that in the Kerr case, for the inclinations $60^\circ$ and
$85^\circ$ the ratio of the observed reflected flux to the observed primary
flux is larger than unity, meaning the reflected component prevails over the
primary one. Because the primary flux is energy-dependent (a power-law), this
ratio depends on energy, too. One can ask what is the maximum and the minimum
value of this ratio over the whole orbit. The energy dependence of this ratio
is shown in the middle right (including the line emission) and right (without
the line) panels in Fig.~\ref{fig-ew}. If the reflected continuum were
without features (Fe lines and the edge), the ratio would
just slightly decrease with energy because of the power-law local primary flux
(see e.g.\/ 5--8~keV and above 10~keV regions for $\theta_{\rm o}=85^\circ$,
top line, in the Schwarzschild case).
The Fe edge causes more complicated behaviour of this
ratio when it influences the minimum (see 8--10~keV region for
$\theta_{\rm o}=85^\circ$, top line, in the Schwarzschild case), or (together
with the flux originating in the iron line) the maximum
(see e.g.\/ 3.8--6.3keV and 4.3--5.2keV region for $\theta_{\rm o}=30^\circ$,
bottom line, in Schwarzschild case).

To evaluate the hardness ratio we compared the fluxes in between 3--6.5 keV
(soft component, $F_{\rm s}$) and 6.5--10 keV (hard component, $F_{\rm h}$).
The hardness ratio is amplified when we compare it with the local hardness
ratio (Fig.~\ref{fig-ratio_loc} and the left panel in
Fig.~\ref{fig-hardness_ratio}).
The amplification is the largest in Kerr case. The value of the $k$-coefficient
in eqs.~(\ref{local_hardness_ratio}) and (\ref{hardness_ratio}) defined by
the primary flux is for these energy ranges $k=0.59$. The local hardness ratio
is larger than this value only by a multiplicative factor 1.01--1.05,
whereas the observed hardness ratio has a maximum larger by $\approx1.2$ in the
Schwarzschild case, and by more than 1.5 in the Kerr case. The sudden increase
in the observed hardness ratio around the time $t/T=0.5$ in Schwarzschild case
and $t/T=0.4$ in Kerr case is due to the fact that the Fe K$\alpha$ line passes
from the soft to the hard
component and back because of the Doppler shift. The flux--flux graph is
shown in the middle panel of Fig.~\ref{fig-hardness_ratio} where the line
with the slope equal to $k=0.59$ is shown as well. Points for all inclinations
are included in this graph. The deviation from this line
is shown in the right panel of the same figure. The points in the flux to flux
graph (for different emission angles) for the local flux lie close to the line
with the slope $\approx 1$.

\section{Summary and conclusions}
\label{conclusions}

We discussed the General Relativity effects in the observed emission of
the spot model. The primary flux was included and the mutual
normalizations were treated within the framework of a simple yet
self-consistent scheme. About half of the isotropic primary flux hits
the spot and is reprocessed (this has been computed by a Monte-Carlo
scheme), and part of the reprocessed radiation is re-emitted towards the
observer. The radiation is influenced by the relativistic effects before
reaching the observer (these have been treated in terms of the transfer
function of the KY code, see Dov\v{c}iak et.~al.\/ \citeyear{dov04c}).
As an example of the two sets of parameters
(a Schwarzschild black hole versus an extremally spinning Kerr black hole)
we find
that all components of the transfer function --- the energy
shifts (Doppler and gravitational), aberration effects (which are in the
interplay with the limb darkening/brightening laws) and the lensing ---
are important and need to be taken into account. As the spot is
orbiting rapidly in the inner regions of the accretion disc, timing is
also essential.  Motion results in the delay amplification, which is
again in a complex  interplay with the bending of light near the black hole;
this is described by eq.~(\ref{kt}).

We would like to emphasize that whereas
the significance of the energy shift and lensing effect
has already been widely studied, the effect
of the finite light travel time has not been discussed much,
in spite of the fact that it significantly affects the resulting signal and
causes an additional enhancement of the observed flux, especially in the case
of a fast-rotating black hole. Although this light time effect arises as an
immediate consequence of the finite velocity of light, its mutual interplay
with the light bending and focusing is rather complicated. We thus examined
this time delay amplification in some detail. Our results also confirm
what has been long suspected, namely, that the local
reprocessing in the disc medium and the signal propagation through the
curved spacetime are all mutually interconnected in a rather complex manner.

We have demonstrated that various integral
characteristics of the spectral features (such as equivalent widths) are at
best moderately sensitive to the black hole rotation.
The equivalent width could be significantly
amplified in our model only if the primary emission were beamed towards the
disc, thus decreasing the observed primary emission.
Both the ratio of the observed reflected to the observed primary flux and the
hardness ratio are amplified when compared to the values for the intrinsic
(local) emission.

In addition to the shown cases, we also performed our computations for
some other values of the spin of the black hole and orbital radius of the spot.
We found out that the results do not
differ significantly for different spin when the radius is kept the same.
We examined the cases with $a=0.998\,GM/c^3$ and $a=0\,GM/c^3$ with
$r=7\,GM/c^2$
and the cases with $a=0.998\,GM/c^3$ and $a=0.93\,GM/c^3$ with $r=3\,GM/c^2$
(the spin in the latter case could not be lower if the spot should be above
the marginally stable orbit). Moreover,
the spectral characteristics depend on the orbital radius mainly close to the
black hole whereas father away their variability either does not change
significantly or slowly decreases.

The main conclusions of this work are summarized as follows.
\vspace*{-\smallskipamount}
\begin{enumerate}
 \item All general and special relativistic effects are important, none of them
   can be neglected.
 \item The EW, apart for the extreme cases of high inclinations,
   does not differ significantly from the local EW,
   however it varies even for low inclination of 30$^\circ$ by up to 30\% when
   compared with its mean value for the whole orbit.
 \item The variability of the reflected to the primary flux ratio and hardness
   ratio changes rapidly with the radius if the spot orbits close to the
   black hole.
 \item The spin of the black hole affects significantly our results only as far
   as it determines the location of the marginally stable orbit.
\end{enumerate}
It follows from the last two points that the flux ratios could be used for
estimating the lower limit of possible values of the spin parameter if the
flare arises in the close vicinity of the black hole.

We must emphasize that nowadays X-ray satellite observatories usually
do not have enough sensitivity to be suitable for real fitting with our
model. However,
%future missions like Xeus or Constellation-X will be
%able to provide data good enough to constrain various parameter values.
future satellite missions like XEUS or CONSTELLATION-X will enable to
spectroscopically follow the motion of individual spots close to the
innermost stable orbit in active galactic nuclei (Goosmann et al.
\citeyear{goo07}).
The azimuthally-dependent effects on the iron line band that we
investigate in this paper can thus be used to constrain the emission
structure and the metric around supermassive black holes.

\section*{Acknowledgments}
This research is supported by the ESA PECS project No.~98040. MD and VK
gratefully acknowledge support from the Czech Science Foundation grants
205/05/P525 and 205/07/0052. GM acknowledges financial support from
Agenzia Spaziale Italiana (ASI). RG is grateful for financial support
to the Centre of Theoretical Astrophysics (LC06014).

{}

\label{lastpage}
\end{document}